\documentclass[twocolumn]{aastex61}
\usepackage{natbib}
\usepackage{array}

\usepackage{graphicx}
\usepackage{latexsym}
\usepackage{amsfonts,amsmath,amssymb}
\usepackage{url}
\usepackage[utf8]{inputenc}
\usepackage{enumitem}
\usepackage{float}
\usepackage{makecell}
\usepackage{xcolor}

\usepackage{placeins}
\hypersetup{colorlinks=true,linkcolor=blue,urlcolor=blue,breaklinks=true,pdfborder={0 0 0},}

\colorlet{Changes@Color}{red}

\newcommand{\dadt}{\left\langle{da/dt}\right\rangle }

\newcommand{\degs}{^\text{$\circ$}}

\newcommand{\aumy}{ \text{ au/My}}
\newcommand{\dadtunit}{ \times 10^{-4} \aumy}
\newcommand{\rate}[2]{$\dadt=(#1\pm#2)\dadtunit$}

\shortauthors{Greenberg et al.}

\turnoffedit

\begin{document}

\title{Yarkovsky Drift Detections for \replaced{159}{247} Near-Earth Asteroids}

\shorttitle{Yarkovsky Drift Detections for 247 Near-Earth Asteroids}

\author{Adam H. Greenberg}
\author{Jean-Luc Margot}
\author{Ashok K. Verma}
\affil{University California, Los Angeles, CA}
\author{Patrick A. Taylor}
\affil{Arecibo Observatory, Universities Space Research Association, Arecibo, PR}
\affil{Lunar and Planetary Institute, Universities Space Research Association, Houston, TX}
\author{Susan E. Hodge}
\affil{Nationwide Children's Hospital, Columbus, OH}

\begin{abstract}

\replaced{The Yarkovsky effect is a thermal process acting upon the
  orbits of small celestial bodies, which can cause these orbits to
  slowly expand or contract with time. The effect is subtle -- typical
  drift rates lie near $10^{-4}$ au/My for a $\sim$1 km diameter
  object -- and is thus generally difficult to measure.  However,
  objects with long observation intervals, as well as objects with
  radar detections, serve as excellent candidates for the observation
  of this effect.  We analyzed both optical and radar astrometry for
  all numbered Near-Earth Asteroids (NEAs), as well as several
  un-numbered NEAs, for the purpose of detecting and quantifying the
  Yarkovsky effect. We present 159 objects with measured drift rates.
  Our Yarkovsky sample is the largest published set of such
  detections, and presents an opportunity to examine the physical
  properties of these NEAs and the Yarkovsky effect in a statistical
  manner. In particular, we confirm the Yarkovsky effect's theoretical
  size dependence of 1/$D$, where $D$ is diameter. We also examine the
  efficiency with which this effect acts on our sample objects and
  find typical efficiencies of around 12\%. We interpret this
  efficiency with respect to the typical spin and thermal properties
  of objects in our sample.  We report the ratio of negative to
  positive drift rates in our sample as $N_{R}/N_{P} = 2.6\pm0.7$ and
  interpret this ratio in terms of retrograde/prograde rotators and
  main belt escape routes.  The observed ratio has a probability of 1
  in 46 million of occurring by chance, which confirms the presence of
  a nongravitational influence.  We examine how the presence of radar
  data affects the strength and precision of our detections. We find
  that, on average, the precision of radar+optical detections improves
  by a factor of approximately 1.6 for each additional apparition with
  ranging data compared to that of optical-only solutions.  }
{The Yarkovsky effect is a thermal process acting upon the orbits of small celestial bodies, which can cause these orbits to slowly expand or contract with time. The effect is subtle ($\dadt \sim 10^{-4}$~au/My for a 1~km diameter object) and is thus generally difficult to measure.  We analyzed both optical and radar astrometry for 600 near-Earth asteroids (NEAs) for the purpose of detecting and quantifying the Yarkovsky effect.  We present 247 NEAs with measured drift rates, which is the largest published set of Yarkovsky detections.  This large sample size provides an opportunity to examine the Yarkovsky effect in a statistical manner.  In particular, we describe two independent population-based tests that verify the measurement of Yarkovsky orbital drift.  First, we provide observational confirmation for the Yarkovsky effect's theoretical size dependence of 1/$D$, where $D$ is diameter.  Second, we find that the observed ratio of negative to positive drift rates in our sample is $2.34$, which, accounting for bias and sampling uncertainty, implies an actual ratio of $2.7^{+0.3}_{-0.7}$.  This ratio has a vanishingly small probability of occurring due to chance or statistical noise.  The observed ratio of retrograde to prograde rotators is two times lower than the ratio expected from numerical predictions from NEA population studies and traditional assumptions about the sense of rotation of NEAs originating from various main belt escape routes.  We also examine the efficiency with which solar energy is converted into orbital energy and find a median efficiency in our sample of 12\%.  We interpret this efficiency in terms of NEA spin and thermal properties.}

\end{abstract}


\section{Introduction}
\label{sec:intro}

The Yarkovsky effect is a small force that results from the
anisotropic thermal emission of small celestial bodies.  Over the past
decade, there has been increasing awareness that the Yarkovsky effect
plays an important role in the evolution of asteroid orbits and the
delivery of meteorites to Earth~\citep{bott06areps}.  Several authors
have published Yarkovsky effect detections for dozens of asteroids:
\citet[][12 detections]{ches08}, \citet[][54 detections]{nuge12yark},
\citet[][47 detections, of which 21 are deemed reliable]{farn13}.
\added{Updates to the latter are given by \citet[][42 valid
  detections]{ches2015} and \citet[][36 valid detections]{vok2015}.}

Here, we provide the largest collection of Yarkovsky detections to
date and introduce several improvements to previous studies.
\citet{nuge12yark} and \citet{farn13} relied on the debiasing of star
catalogs proposed by \citet{ches10}.  Our current model uses the more
up-to-date and accurate debiasing algorithm of
\citet{farn15}. Previous works have traditionally relied on a
signal-to-noise (S/N) metric and the quantity and quality of the
observational data to distinguish between detections and
nondetections~\citep{ches08,farn13}, or by augmenting these criteria
with an explicit sensitivity metric \citep{nuge12yark}.  Here, we
further refine the detection criterion with a precise formulation
based on an analysis of variance~\citep{gree17}.  Some of the previous
formulations \citep[e.g.,][]{nuge12yark} included a finite increment
in semi-major axis at each time step irrespective of the asteroid's
distance from the Sun.  Here, we use a 1/$r^2$ dependence of the solar
flux.  The \citet{nuge12yark} results were based on astrometry
obtained as of January 31, 2012.  The current work benefits from more
than \replaced{5}{7} years of additional astrometry, including more
than \replaced{100}{250} additional ranging observations with the
Arecibo and Goldstone radars.  Finally, the numbers of known NEAs and
numbered NEAs have both \replaced{nearly}{more than} doubled since the
\citet{nuge12yark} study.  The number of detections is now
sufficiently large that ensemble properties can be refined, such as
the ratio of retrograde to prograde rotators, and the physical theory
can be tested, such as the dependence of the Yarkovsky drift magnitude
as a function of asteroid size.

\section{Data preparation}
\label{sec:dataprep}

Optical astrometry was automatically downloaded from the Minor Planet
Center (MPC) on \replaced{Mar 8, 2017}{Nov 11,
  2019}~\citep{mpcdb}. Astrometry taken from nonstationary
(generally, space-based) observatories was
discarded.
\added{The number of optical
  {\em observations} considered in this work is 379,434.  Each optical
  observation yields two {\em measurements} of position on the plane
  of the sky.}  Radar astrometry was downloaded from the JPL Radar
Astrometry Database \citep{radardb} and was discarded from MPC records
to avoid duplication.  In a few instances, previously unpublished
radar data obtained by the authors were also used.  \added{The radar
  data considered in this work include 735 range measurements and 412
  Doppler measurements.}

\subsection{Weighting and debiasing}
\label{sec:weighting}

Optical astrometry was weighted following the methods described by
\citet{farn15}. To summarize, this method involved weighting
measurements based on the observatory, type of measurement, star
catalog, and date. We also used the ``batched weighting'' scheme
described by \citet{farn15}, wherein measurements taken from
the same observatory on the same night were given a smaller
weight. Star catalog debiasing was also performed according to the
approach of \citet{farn15}.

Radar astrometry was weighted according to observer-reported
uncertainties.

\subsection{Outlier rejection}
\label{sec:outlierrejection}

Outlier rejection was performed via an iterative fit-drop-add
scheme. All available data were used during the initial gravity-only
orbital fit. Then, all
optical measurements with weighted residuals beyond a fiducial
threshold were rejected.
\deleted{Radar data were excluded from outlier rejection.}
This threshold was defined as
\begin{equation}
\frac{(O_{i,RA}-C_{i,RA})^2}{\sigma_{i,RA}^2} + \frac{(O_{i,DEC}-C_{i,DEC})^2}{\sigma_{i,DEC}^2}  < 8,
\end{equation}
where $O$ and $C$ stand for observed and computed values,
respectively, RA and DEC stand for right ascension and declination,
respectively, $\sigma$ represents observational uncertainty, and the
index $i$ represents the $i^{th}$ observation.

As the fit iterated, previously discarded measurements were
reevaluated with respect to this threshold, and included in
subsequent iterations, as appropriate.
Outlier rejection was disabled after three fit-drop-add iterations gave identical results.

Initially, outlier rejection was performed with a gravity-only
model. After the
Yarkovsky component of the dynamical model was estimated, outlier
rejection was performed once more with the additional
Yarkovsky 
component included (Section~\ref{sec:forcemodel}).

\added{For the objects analyzed in this work, the median percentage of
  rejected observations was 1.6\% of the total number of observations,
  with a standard deviation of 1.6\% of the total number of
  observations. The largest percentage of rejected observations in our
  sample did not exceed 10\%, with the exception
    of (408982) 2002~SP for which 18 observations out of 100 were
    rejected, almost all of which came from the same observatory.}

\added{Because radar measurement residuals are typically superb, radar
  data were excluded from outlier rejection.  The mean and standard
  deviation of the radar residuals normalized to the reported
  uncertainties of the measurements are $0.30$ and $0.33$,
  respectively.  In our entire data set, the three largest normalized
  range residuals are 1.32, 1.65, and 2.04.}

\section{ Orbit determination }
\label{sec:idos}

Orbit determination was performed using our Integration and Determination of
Orbits System (IDOS, see \citet{gree17}). At its core, this software utilizes the Mission analysis,
Operations, and Navigation Toolkit Environment (MONTE), a set of tools developed
by the Jet Propulsion Laboratory (JPL) for a variety of space-related science
and aeronautical goals \replaced{\citep{evans16}}{\citep{evan18}}.  The MONTE orbital integrator can
account for gravitational perturbations from any set of masses -- for the
analyses performed in this paper, we considered the
eight known planets and 24 of the
most massive minor planets~\citep{folkner2014} as gravitational perturbers.
During close Earth approaches, the integrator considers a detailed model of the
planetary gravitational field. MONTE also accounts for general
relativistic effects during orbital integration.
Further details concerning the
internal operations of IDOS
were described by \citet{gree17}.

In gravity-only solutions, we estimated
the six parameters
(three position and three velocity components)
of the state vector 
simultaneously.  In
Yarkovsky solutions, we estimated an additional parameter describing
the strength of the Yarkovsky drift. 
We assigned one-standard-deviation uncertainties
($\sigma$) to our Yarkovsky estimates such that a 1-$\sigma$ change to
the drift rate results in an increase of one in the sum of squares of
weighted residuals, similar to the approach of \citet{nuge12yark}.
This approach yields values that match the formal uncertainties
derived from a covariance matrix, which was the approach of
\citet{farn13}.

\section{ Yarkovsky force model }
\label{sec:forcemodel}

We utilized the Yarkovsky force model described
by \cite{gree17}, where the magnitude and direction of the
thermal acceleration, $\ddot{r}$, are calculated and applied at every
integration time step of the dynamical model. The acceleration is
calculated as
\begin{equation} \label{eqn:ayark}
	\ddot{\vec{r}}= \zeta \; \frac{3}{8\pi} \; \frac{1}{D\rho} \; \frac{L_{\odot}}{c} \; \frac{X_{\hat{p}}(\phi)\vec{r}(t)}{||\vec{r}(t)||^3},
\end{equation}
where $\vec{r}(t)$ is the heliocentric radial vector for the object at time $t$,
$\hat{p}$ is the unit spin-axis vector, $\phi$ is the phase lag,
$L_{\odot}$ is the luminosity of the Sun, $c$ is the speed of light,
and $X_{\hat{p}}(\phi)$ is the rotation matrix about \replaced{$p$}{$\hat{p}$}. $D$ and $\rho$ are the
diameter and density of the object, respectively, while $\zeta$ is an
efficiency factor.
The phase lag $\phi$ describes the longitude on the surface from which photons are reemitted, relative 
to the sub-Solar longitude.
In Equation (\ref{eqn:ayark}), we assume a perfect absorber, i.e., a
Bond albedo of zero.

For the objects analyzed in this work, specific values for $\phi$ and
$\hat{p}$ were not known. Therefore, these values were fixed at
$90\degs$ and antiparallel to the orbit normal vector, respectively
, which maximizes the magnitude of the orbital perturbation.
As we discuss in the \replaced{next paragraph}{following paragraphs}, these assumptions do not affect the
estimated value of the semi-major axis drift.

\added{We also note that this treatment models the diurnal Yarkovsky
  effect, but not the seasonal effect. The seasonal effect, which is
  caused by the apparent rotation of an object orbiting the Sun, tends
  to be small compared to the diurnal effect
  \citep{vokrouhlicky2000}. A complete formulation accounting for both
  the seasonal and diurnal effects is described in \citet{vok2015}.}

With knowledge of the orbit semi-major axis, $a$, and eccentricity, $e$,
the orbit-averaged drift in semi-major axis, $\dadt$, 
can be determined from this acceleration model with
\begin{equation}
\label{eqn:yarkfundamental}
  \dadt = \pm\;\xi \frac{3}{4\pi} \frac{1}{\sqrt{a}} \frac{1}{1-e^2} \frac{L_{\odot}}{c\sqrt{GM_{\odot}}} \frac{1}{D\rho},
\end{equation}
which is equivalent to \citet{gree17}'s equation (8) and
corrects \citet{nuge12yark}'s equation (1).
Here, $\xi$ is the Yarkovsky efficiency, and depends on $\zeta$, spin pole
obliquity $\gamma$
(i.e., the angle between the spin pole vector $\hat{p}$ and the orbit
normal vector), and phase lag $\phi$. We always take the Yarkovsky
efficiency to be positive.  Any incorrect assumption about Bond
albedo, diameter, \added{density,} obliquity, and phase lag is absorbed in this
efficiency factor such that the $\dadt$ value, which is dictated by
the astrometry, is not affected by these assumptions
(Section~\ref{sec:interpretxi}).

With numerical values, we find
\replaced{\begin{equation}
\label{eqn:yarkfinal}
  \begin{aligned}
\dadt =  \pm1.45 \left( \frac{\xi}{0.01} \right) \left( \frac{1 \text{ au}}{a}\right)^{\frac{1}{2}} \left( \frac{1}{1-e^2} \right) \; \times\\
	\left( \frac{1 \text{ km}}{D} \right) \left( \frac{ 1000 \text{ kg}\text{ m}^{-3}}{\rho} \right)    
        \times  \frac{10^{-4}\text{ au}}{\text{My}}.
  \end{aligned}
\end{equation}}{\begin{equation}
\label{eqn:yarkfinal}
  \begin{aligned}
    \dadt =  \pm14.4 \left( \frac{\xi}{0.1} \right) \left( \frac{1 \text{ au}}{a}\right)^{\frac{1}{2}} \left( \frac{1}{1-e^2} \right) \; \times\\
	\left( \frac{1 \text{ km}}{D} \right) \left( \frac{ 1000 \text{ kg}\text{ m}^{-3}}{\rho} \right)    
    \times  \frac{10^{-4}\text{ au}}{\text{My}}.
  \end{aligned}
\end{equation}
}

\section{ Candidate selection }
\label{sec:selection}

\subsection{ Initial selection}
\label{sec:initialselection}

We considered \replaced{three}{four} sets of Yarkovsky detection candidates.
Two sets of candidates, the \texttt{Nugent12} set and the
\texttt{Farnocchia13} set,
represent 
Yarkovsky detections
reported by \citet{nuge12yark} and \citet{farn13},
respectively. For these objects, we performed our analysis in two ways
-- first, by using the same observational data as those used by the
authors, and second, by using all currently available data
(Section~\ref{sec:refinement}).  The \texttt{Nugent12} set features 54
objects, while the \texttt{Farnocchia13} set contains 47 objects.

The third \deleted{data} set\deleted{, \texttt{UCLA17},} contains
objects that had not previously been considered by the other two works
but that we determined to be Yarkovsky detection candidates. For the
most part, these objects had either not yet been discovered, or had
small observation intervals prior to 2012 or early 2013.  We
identified the new candidates as follows.  First, we downloaded the
list of \replaced{15,595}{21,135} known NEAs from the MPC on
\replaced{March 7, 2017}{November 11, 2019}.  Second, for each one of
the \replaced{2,348}{2915} numbered NEAs, we computed the Yarkovsky
sensitivity metric ($s_Y$) described by \citet{nuge12yark}.  This
root-mean-square quantity provides an \deleted{excellent} assessment of the
relative sensitivity of selected data sets to drifts in semimajor axis
\added{on the basis of optical astrometry}.  We used the threshold
determined by \citet{nuge12yark} of $s_Y \geq 2$.
\replaced{Only 376}{We found that 567} NEAs met this condition.
\deleted{ Third, we computed preliminary estimates of $\dadt$ and
  associated uncertainties for these 376 NEAs.  We defined a
  signal-to-noise (S/N) metric as the ratio of the best-fit $\dadt$ to
  its one-standard-deviation uncertainty.  We selected the 200 NEAs
  that have both $s_Y > 2$ and S/N $>$ 1.}  \deleted{Of these, 59 had
  been previously considered by \citet{nuge12yark} or
  \citet{farn13}.}

\added{The fourth and final set includes 24 additional objects of
  particular interest, including 22 numbered binary asteroids.
  Yarkovsky detections of binary asteroids are important because the
  masses and obliquities of the binaries are measurable and frequently
  known~\citep[e.g.,][]{marg15AIV}, enabling a direct interpretation of
  the Yarkovsky drift rate in terms of asteroid thermal properties
  \citep[e.g.,][]{marg04,vokr05bin}.  We also included 137924 (2000 BD19) to
  ensure a complete sampling of objects that are actively being
  tracked to test general relativity and measure the oblateness of the
  Sun~\citep{marg09iau261,verm17gr}. Measuring the Yarkovsky drift
  rates of these objects whose perihelion longitudes precess rapidly
  is important to recognize and quantify the various dynamical
  influences affecting their trajectories.  Finally, we include 441987
  (2010 NY65), which exhibits a horseshoe orbit similar to that of
  54509 YORP~\citep{lowr07,tayl07} and presents repeated opportunities
  for high-precision dynamical work with radar.}

\added{Among the four sets of objects, there are 600 distinct
  Yarkovsky candidates.}

\citet{nuge12yark} rejected Yarkovsky detections for which there were
fewer than 100 astrometric measurements, or for which the observation
interval was less than 15 years. However, we reviewed the detections
that were discarded due to these criteria in 2012 and found that 90\%
of them \replaced{were}{are} reliable, i.e., their $\dadt$ values are
consistent with values presented in this work, even after the addition
of post-2012 data.  In this work, we flag objects that
\citet{nuge12yark} would have discarded because of data span or
quantity, but we do not discard the detections\added{, unless the
  observation interval is shorter than 5 years}.

\deleted{Among the three sets of objects (\texttt{Nugent12},
\texttt{Farnocchia13}, and \texttt{UCLA17}), there
are 231 distinct Yarkovsky candidates.}

\subsection{Selection refinement}
\label{sec:refinement}

After candidate selection, we performed a six-parameter fit to the
astrometry using a gravity-only model, followed by a seven-parameter
fit which included a Yarkovsky force model.  We then performed an {\em
  analysis of variance} \citep{greenbook} to determine whether the data warrant the use
of the Yarkovsky model.

Specifically, we calculated the test statistic
\begin{equation}
F = \frac{ \kappa_{\delta} }{ \kappa_{Y} }
\end{equation}
where 
\begin{equation}
\kappa_{\delta} =	\frac{ \sum_{i=0}^{N} (\frac{O_i - C_{0,i}}{\sigma_i})^2  -
			       \sum_{i=0}^{N} (\frac{O_i - C_{Y,i}}{\sigma_i})^2 }{m_Y-m_0} 
\end{equation}
and
\begin{equation}
  \kappa_{Y} = \frac{\sum_{i=0}^{N} (\frac{O_i - C_{Y,i}}{\sigma_i})^2 }{N - m_Y}.
\end{equation}
Here, $C_{0,i}$ is the $i^{th}$ computed value assuming gravity only,
$C_{Y,i}$ is the $i^{th}$ computed value assuming our best-fit
Yarkovsky model, $O_i$ is the $i^{th}$
\replaced{observation}{measurement} and $\sigma_i$ is the
\replaced{measurement uncertainty for that observation}{associated
  uncertainty}, $N$ is the number of
\replaced{observations}{measurements}, and $m_Y$, $m_0$ are the number
of free parameters in the Yarkovsky model ($m_Y=7$) and gravity-only
model ($m_0=6$), respectively.

We then calculated the value 
\begin{equation}
p = \int_{x=F}^{x=\infty} f(m_Y-m_0, N-m_Y, x)dx ,
\end{equation}
where $f(m_Y-m_0, N-m_Y, x)$ is the F-distribution probability density
function with $m_Y-m_0$ and $N-m_Y$ degrees of freedom.  The $p$-value
serves as a metric for testing the null hypothesis --- namely, that
the additional degree of freedom introduced by the Yarkovsky force
model is superfluous.

Our initial selection refinement step consisted of discarding those
objects for which $p > 0.05$, which approximately corresponds to a
two-standard-deviation detection threshold. This step rejected
\replaced{60}{283} objects, leaving \replaced{171}{317} objects for
further consideration.

\replaced{We also followed the procedure of \citet{nuge12yark}, and
  determined those objects for which there were fewer than 10
  measurements in the first 10 years of observations.}{We also
  implemented a robustness test where we eliminated the 10 earliest
  observations from the optical astrometry of each remaining object.}
\deleted{This check is necessary because isolated, erroneous
  astrometry can result in spurious detections.}  For these objects,
we refit the Yarkovsky model with the \replaced{sparse}{early}
observations removed, and rejected any object\replaced{s for which the
  resulting $\dadt$ value changed significantly from that of the
  nominal fit}{ that no longer met the $p \leq 0.05$ criterion.
  Objects were also rejected when $p$ was $\leq 0.05$, but the error
  bars of the Yarkovsky rates with and without the early observations
  did not overlap.}  \added{This check is necessary because early
  astrometry, which can be of lower quality or erroneous, often yields
  spurious detections.}  This step rejected \replaced{12}{60} objects,
leaving \replaced{159}{257} objects \replaced{remaining}{for further
  analysis}.
\deleted{These 159 asteroids make
  up our final set of Yarkovsky detections (Table~\ref{tbl:data}).}

Finally, because pre-CCD astrometry can lead to spurious detections
(Section \ref{sec-ganymed}) even with proper weights, we
reanalyzed \replaced{27}{24}
\added{remaining} Yarkovsky candidates
for which pre-1965 astrometry exists.  Specifically, we discarded the
pre-1965 astrometry, fit for $\dadt$ values with the shortened
observation intervals, and recomputed $p$-values.  Objects that no
longer met the $p \leq 0.05$ criterion were flagged.  \added{Objects
  were also flagged when $p$ was $\leq 0.05$, but the error bars of the
  Yarkovsky rates with and without the pre-1965 observations did not
  overlap.}  \replaced{About a dozen}{Eight} objects
\replaced{are in
  this category}{failed this test} and their Yarkovsky rates require
additional verification.

\added{At the end of this process, 249 objects remained in our data
  set.  Two objects had arc lengths shorter than 5 years and were
  eliminated. The 247 remaining asteroids constitute our final set of
  Yarkovsky detections.  They include 122 Apollos (49\%), 81 Atens
  (33\%), and 44 Amors (18\%).  With the exception of 2009~BD, they
  span a range of absolute magnitudes between 12.4 and 24.4.  Our
  Yarkovsky drift rate and efficiency measurements are shown in
  Table~\ref{tbl:data}.  For completeness, we also list the 8 objects
  whose rates differ when including or excluding pre-1965 observations
  in Table~\ref{tbl:pre1965}.}

\section{ Comparison with previous works }
\label{sec:comparison}

\replaced{Many}{Approximately 25\%} of the objects
\replaced{considered for}{reported in} this work had been previously
reported as Yarkovsky detections (Section~\ref{sec:selection}). It is
useful to compare our Yarkovsky determinations to these previous
works, for two reasons. First, because our results were determined
independently of the previous works, a comparison serves as a check on
both sets of results.  Second, new astrometry has been reported for
many of these objects.  Therefore, we can study how the results and
uncertainties changed in light of new data.

We performed two comparisons with the previous works.
In each case, we compared both our absolute Yarkovsky measurements and
their associated uncertainties to those of the original works. 
We first created data sets that roughly matched the observational
intervals reported by previous authors, to the nearest calendar year.
In doing so, we expect there to be good agreement between our
Yarkovsky detections and those of the original works.  We do
anticipate slight differences introduced by our use of improved
debiasing and weighting algorithms (Section~\ref{sec:weighting}) and
by our use of observation sets that are not identical to those used in
the original works
(e.g., observations at beginning or end of
intervals matched to the nearest calendar year, precovery
observations, or observations that were remeasured).  For our second
comparison, we included all available data for all objects.  In this
case, we expect an overall lower level of agreement because of our use
of additional astrometry, which sometimes represent a significant
fraction of the available astrometry.

Because we are interested in whether our results match those
previously published, it is useful to quantify what we mean by a
``match''.  We used a metric inspired by mean-comparison
tests. Namely, for each object $i$ in the dataset, we calculated
\begin{equation}
	z_i = \frac{|Y_{t,i} - Y_{p,i}|}{\sqrt{\sigma_{t,i}^2 + \sigma_{p,i}^2}},
\end{equation}
where $Y_{t,i},Y_{p,i}$ are this work's estimated drift rate for
object $i$ and the previous work's estimated drift rate for object $i$,
respectively, and
$\sigma_{t,i},\sigma_{p,i}$ are this work's uncertainty for object $i$
and the previous work's uncertainty for object $i$, respectively. The
quantity $z$ therefore represents a significance score. By choosing a
threshold value for $z$, we can
signal our confidence that our
measurement is consistent with that of the original work. We chose a
significance threshold of $2.0$, i.e.,
detection $i$ was considered a match if
\begin{equation}
	z_i < 2.0.
\end{equation}
In other words, we concluded that the two measurements matched if we
could not reject the hypothesis that the two measurements were drawn
from the same distribution at the 95\% confidence level.

\section{Yarkovsky drift rates}

We measured semi-major axis drift rates and calculated Yarkovsky efficiency
values for \replaced{159}{247} NEAs,
shown in Table~\ref{tbl:data} and ordered by object number. We present
drift rates derived from optical measurements, as well as optical plus radar
astrometry.
A machine-readable file containing the data in this table can be
found at
\href{http://escholarship.org/uc/item/0pj991hd}{http://escholarship.org/uc/item/0pj991hd}.

\startlongtable
\begin{deluxetable*}{lrrrrrrrrrrrrrrl}
\tabletypesize{\scriptsize} 
\tablehead{
\colhead{Name} & \colhead{ $a$ } & \colhead{ $e$}  & \colhead{$D$} & \colhead{ $N_{o}$} & \colhead{$N_{r}$} & \colhead{$\dadt$} & \colhead{$\sigma$} & \colhead{ $p$} & \colhead{$\dadt_r$} & \colhead{$\sigma_r$} & \colhead{$p_r$} & \colhead{$s_Y$} & \colhead{$\xi$} & \colhead{Arc }
} \tablecaption{Yarkovsky drift measurements for
  \protect\replaced{159}{247} Near-Earth Asteroids.  The semi-major
  axis, $a$, is in au. $e$ is orbital eccentricity. Asteroid diameter
  $D$ is in km. \protect\added{Approximately half of the diameter
    values were extracted from JPL's Small Body Database (SBDB)}.
  Diameters inferred from $H$ magnitude via Equation (\ref{eqn:diam}) are
  flagged with a \protect\added{$T$ when the taxonomic type is
    available from the SBDB and with a }* \protect\added{otherwise}.  $N_{o}$,
  $N_{r}$ are the number of optical measurements and radar
  measurements \protect\added{used in the solution}, respectively.
  $\dadt$, $\sigma$, the semi-major axis drift and associated
  uncertainty, are in $\dadtunit$.  For objects with radar astrometry
  ($N_{r}>0$), we also report $\dadt_r$ and $\sigma_r$, which
  incorporate those radar measurements. $p$ and $p_r$ are the p-values
  used in distinguishing between a gravity-only dynamical model and a
  Yarkovsky dynamical model using optical data only and optical plus
  radar data, respectively. $s_Y$ is the Yarkovsky sensitivity
  parameter of \citet{nuge12yark}.  $\xi$ indicates the Yarkovsky
  efficiency, which was computed with a bulk density that was
  \protect\added{extracted from the SBDB, if available, or} inferred
  from the spectral type, if available (Section~\ref{sec:xi}).
  Yarkovsky detections \protect\replaced{determined}{predicted} to be
  weaker because of the time span or quantity of astrometry
  (Section~\ref{sec:initialselection}) are flagged with $^{\dagger}$,
  whereas objects with anomalously high $\xi$ values ($\xi>0.5$) are
  flagged with $^{\mathsection}$.  \protect\deleted{Objects that yield
    a drift rate detection with pre-1965 data but not without it
    (Section~\ref{sec:refinement}) are flagged with
    $^\text{C}$.}\protect\added{Binary and triple asteroids are
    flagged with $^B$.}  \protect\authorcomment1{Numerical values in
    this table have been updated.}
\label{tbl:data}
}
\startdata
\hline
(1566) Icarus &  1.08  &  0.83  &       1.00   &  2372  & 16  &    -4.47  &      0.4  & 1e-16  &    -4.84  &      0.4  &    1e-16  &     37.1  &     0.03  & $1949-2019$ \\
                       (1620) Geographos &  1.25  &  0.34  &       2.56   &  8890  &  7  &    -0.68  &      0.5  & 3e-03  &    -1.02  &      0.5  &    2e-06  &     18.7  &     0.05  & $1951-2019$ \\
                             (1685) Toro &  1.37  &  0.44  &       3.40   &  5050  &  9  &    -1.52  &      0.4  & 1e-16  &    -1.57  &      0.4  &    1e-16  &     33.0  &     0.10  & $1948-2019$ \\
        (1864) Daedalus$^{\mathsection}$ &  1.46  &  0.61  &       3.70   &  4368  &  1  &   -11.29  &      5.9  & 6e-05  &   -11.67  &      5.8  &    3e-05  &      3.0  &     0.56  & $1971-2019$ \\
                         (1865) Cerberus &  1.08  &  0.47  &       1.20   &  3396  &  0  &    -3.75  &      1.8  & 3e-06  &      $-$  &      $-$  &      $-$  &     11.5  &     0.07  & $1971-2019$ \\
                   (1866) Sisyphus$^{B}$ &  1.89  &  0.54  &       8.48   &  7796  &  1  &    -2.26  &      2.5  & 2e-02  &    -2.26  &      2.5  &    2e-02  &      4.3  &     0.35  & $1955-2019$ \\
                           (1916) Boreas &  2.27  &  0.45  &       3.50   &  2892  &  0  &    -4.86  &      2.4  & 7e-06  &      $-$  &      $-$  &      $-$  &      7.7  &     0.39  & $1953-2019$ \\
                             (2062) Aten &  0.97  &  0.18  &       1.10   &  1922  &  7  &    -6.06  &      0.9  & 1e-16  &    -5.34  &      0.7  &    1e-16  &     24.1  &     0.10  & $1955-2019$ \\
\enddata
\end{deluxetable*}

\startlongtable
\begin{deluxetable*}{lrrrrrrrrrrrrrrl}
\tabletypesize{\scriptsize} 
\tablehead{
\colhead{Name} & \colhead{ $a$ } & \colhead{ $e$}  & \colhead{$D$} & \colhead{ $N_{o}$} & \colhead{$N_{r}$} & \colhead{$\dadt$} & \colhead{$\sigma$} & \colhead{ $p$} & \colhead{$\dadt_r$} & \colhead{$\sigma_r$} & \colhead{$p_r$} & \colhead{$s_Y$} & \colhead{$\xi$} & \colhead{Arc }
} \tablecaption{Yarkovsky drift measurements for 8 Near-Earth
  Asteroids whose rates require additional verifications because
  solutions that include pre-1965 astrometry, shown here, differ from
  solutions that exclude pre-1965 astrometry.  Columns as in Table~\ref{tbl:data}.
  \protect\authorcomment1{This is a new table.}
\label{tbl:pre1965}
}
\startdata
\hline
(433) Eros$^{\dagger}$ &  1.46  &  0.22  &       16.84  & 16850  &  5  &    -0.49  &      0.2  & 1e-13  &    -0.43  &      0.1  &    2e-13  &     66.3  &     0.15  & $1900-2019$ \\
                (719) Albert$^{\dagger}$ &  2.64  &  0.55  & $^{T}$2.38   &  3604  &  0  &    -1.83  &      1.4  & 3e-04  &      $-$  &      $-$  &      $-$  &     11.9  &     0.09  & $1911-2019$ \\
(1036) Ganymed$^{\dagger}$$^{\mathsection}$ &  2.66  &  0.53  &       37.67  & 12028  &  1  &    -4.98  &      1.3  & 1e-16  &    -4.98  &      1.3  &    1e-16  &      6.4  &     4.14  & $1924-2019$ \\
         (1862) Apollo$^{B}$$^{\dagger}$ &  1.47  &  0.56  &       1.50   &  3216  & 17  &    -0.30  &      0.9  & 5e-01  &    -1.78  &      0.2  &    1e-16  &     28.9  &     0.04  & $1930-2019$ \\
         (1915) Quetzalcoatl$^{\dagger}$ &  2.54  &  0.57  &       0.50   &    96  &  1  &   -42.38  &     23.0  & 3e-02  &   -45.18  &     22.7  &    2e-02  &      4.0  &     0.41  & $1953-2004$ \\
                 (2061) Anza$^{\dagger}$ &  2.26  &  0.54  &       2.60   &  1822  &  0  &     3.29  &      3.8  & 5e-02  &      $-$  &      $-$  &      $-$  &      6.2  &     0.16  & $1960-2019$ \\
(4015) Wilson-Harrington$^{\dagger}$$^{\mathsection}$ &  2.63  &  0.63  &       4.00   &  1874  &  0  &    -7.57  &      3.1  & 1e-04  &      $-$  &      $-$  &      $-$  &     10.3  &     0.51  & $1949-2019$ \\
             (29075) 1950 DA$^{\dagger}$ &  1.70  &  0.51  &       2.00   &  1180  & 12  &    -3.92  &      1.8  & 2e-05  &    -2.61  &      0.5  &    1e-16  &     13.3  &     0.09  & $1950-2019$ \\
\enddata

\end{deluxetable*}

\section{ Comparison with Nugent et al. (2012)}
\label{sec:nugent}

\subsection{ Using matching observation intervals}

\begin{figure*}[p!]
\centering
\includegraphics[width=1.95\columnwidth]{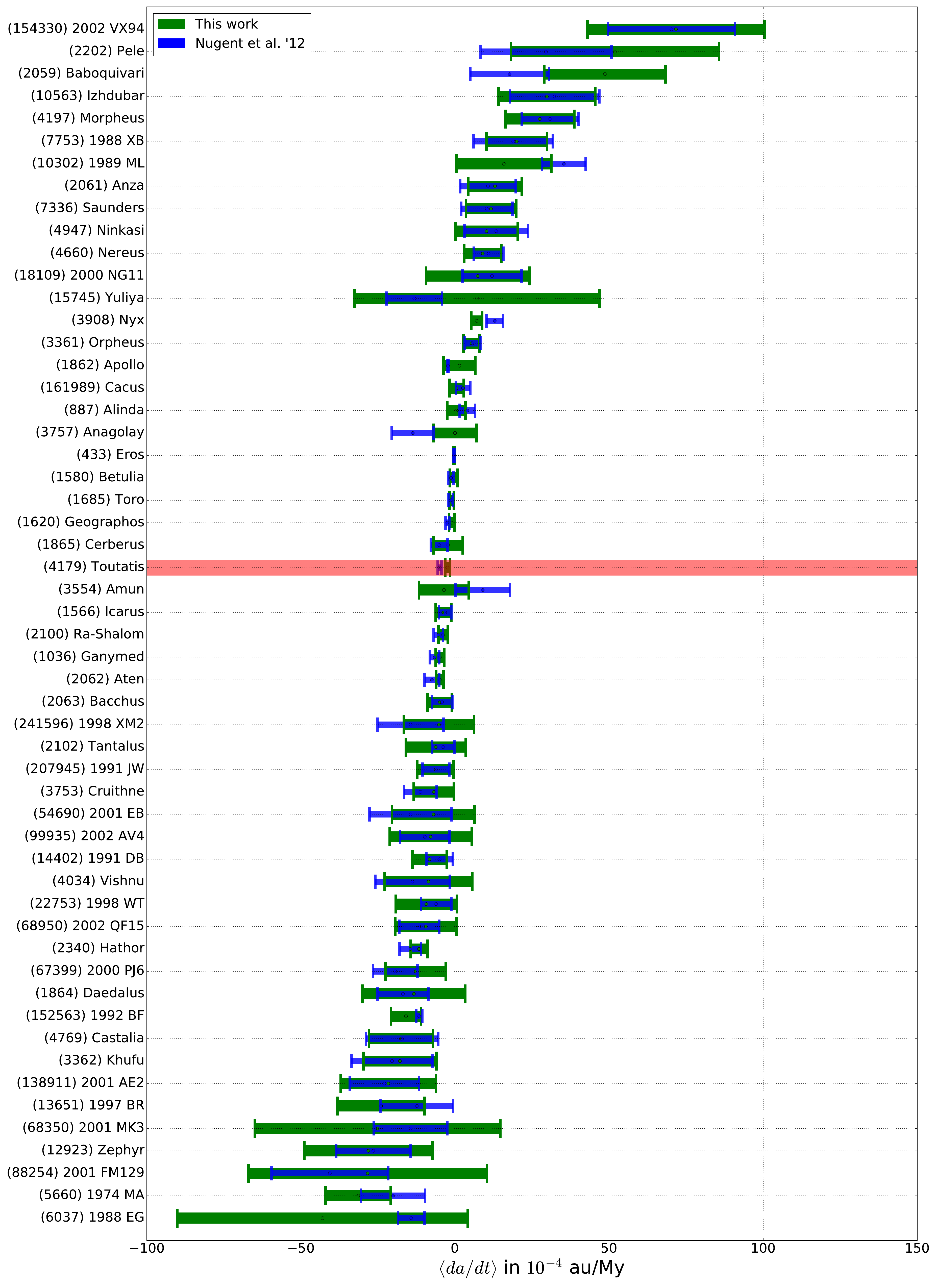}
\caption{A comparison of our Yarkovsky detections (green) and those determined by
  \citet{nuge12yark} (blue),
  when we used only matching data.
  Measurements that disagreed (i.e., $z_i > 2$, Section~\ref{sec:comparison})
  are highlighted in red.
  Objects are ranked from most positive to most negative Yarkovsky drift rate.
}
\label{fig:nugentmatch}
\end{figure*}

We analyzed the 54 Yarkovsky objects described by \citet{nuge12yark}
by
constructing observation intervals whose calendar years matched those
listed in Table 3 of that work.  We compared
(Section~\ref{sec:comparison}) our results with their findings
(Figure~\ref{fig:nugentmatch}).
We agreed with all $\dadt$ values save one, (4179) Toutatis, for which
we found a $z$-score of 2.68.  We examine this object in more detail
in Section~\ref{sec-Toutatis}.

However, we also found that \replaced{23}{16} objects that
\citet{nuge12yark} identified as detections did not pass our detection
threshold (Section~\ref{sec:refinement}).  Much of this discrepancy is
explained by this work's higher threshold for detection --- a
$p$-value of 0.05 approximately corresponds to an S/N of 2, while
\citet{nuge12yark} considered possible detections for objects with S/N
$>$ 1.  Indeed, all but \replaced{five}{two} of the \replaced{23}{16}
objects exhibit 1 $<$ S/N $<$ 2 in \citet{nuge12yark}'s table.

\subsection{ Using all available data}

\begin{figure*}[p!]
\centering
\includegraphics[width=1.95\columnwidth]{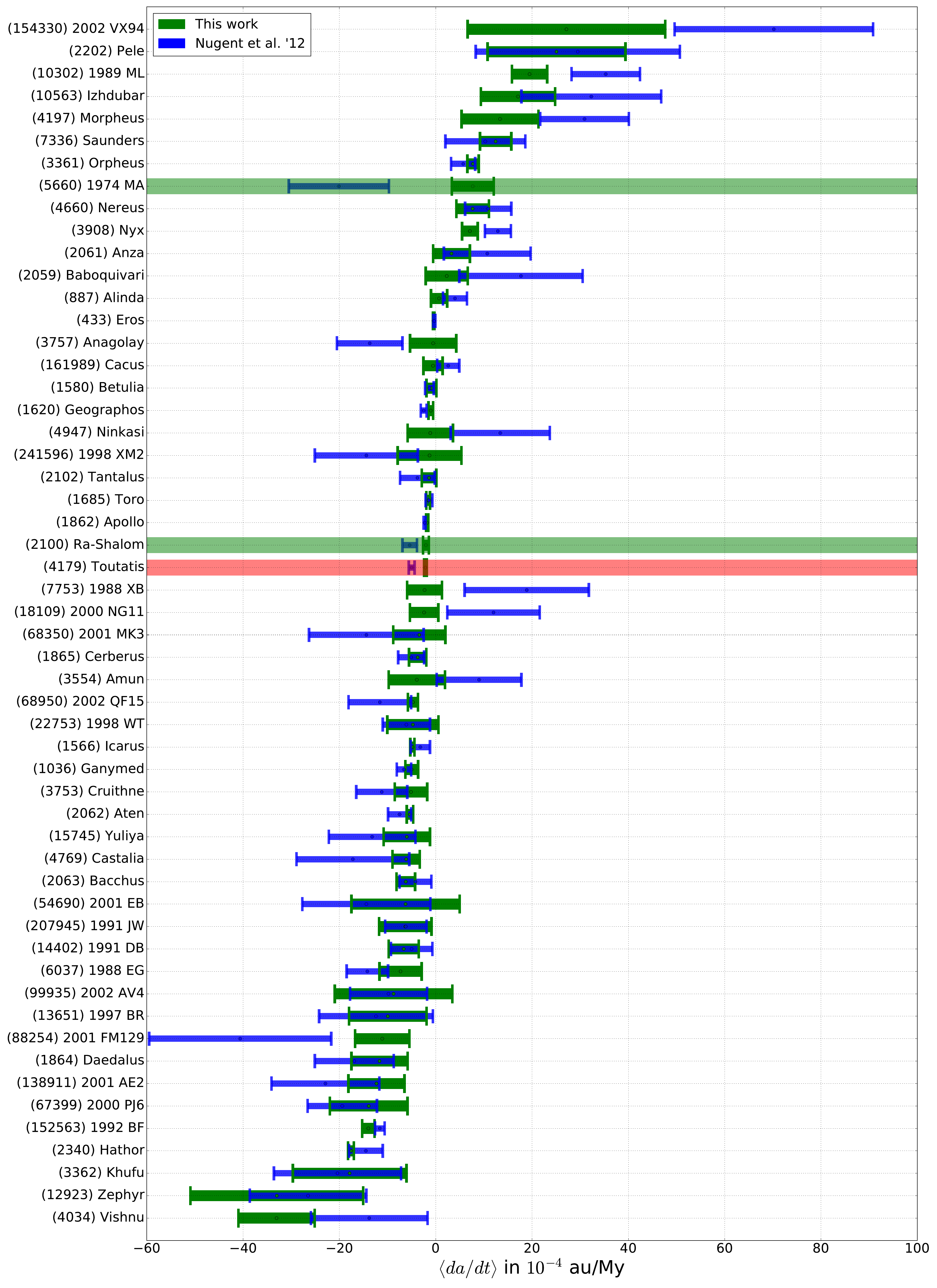}
\caption{A comparison of our Yarkovsky detections (green) and those determined by
  \citet{nuge12yark} (blue), when we used all available data.
  Measurements that disagreed (i.e., $z_i > 2$, Section~\ref{sec:comparison}) 
  only when using all available data are highlighted in green, while those that
  also disagreed
  when matching observational intervals are highlighted in red.
  Objects are ranked from most positive to most negative Yarkovsky drift rate.}
\label{fig:nugent}
\end{figure*}

When using all available data
(including data that were not available for use by
\citet{nuge12yark}), 
we found good agreement (Figure~\ref{fig:nugent}), except for \replaced{two}{three}
objects --- \added{(2100) Ra-Shalom, }(4179) Toutatis\replaced{ and (1620) Geographos}{, and (5660) 1974 MA} --- for which our
drift rates do not match those of \citet{nuge12yark}.  \added{We discuss these special cases in Sections \ref{sec-Toutatis} and \ref{sec:farnnomatch}.}

\begin{figure*}[p!]
\centering
\includegraphics[width=1.95\columnwidth]{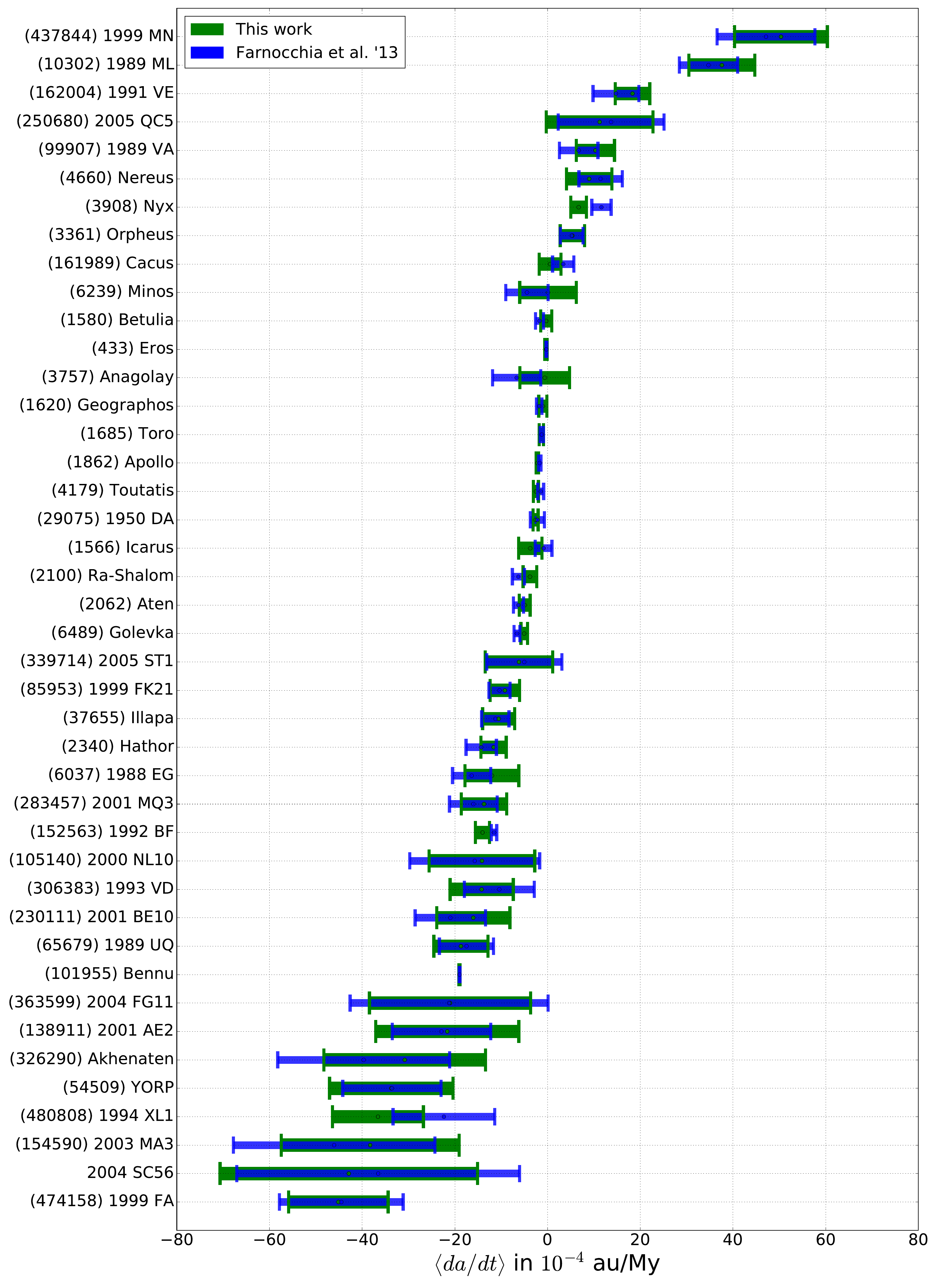}
\caption{A comparison of our Yarkovsky detections (green) and those
  determined by \citet{farn13} (blue), when we used matching
  data arcs.
  Some objects
  ((483656) 2005 ES70, 2003 XV, 2004 BG41, 2007 PB8, 2009 BD) were not
  included in this plot for display purposes, but all of them had $z_i <= 2$.
  Objects are ranked from most positive to most negative Yarkovsky drift rate.
}
\label{fig:farnmatch}
\end{figure*}

\begin{figure*}[p!]
\centering
\includegraphics[width=1.95\columnwidth]{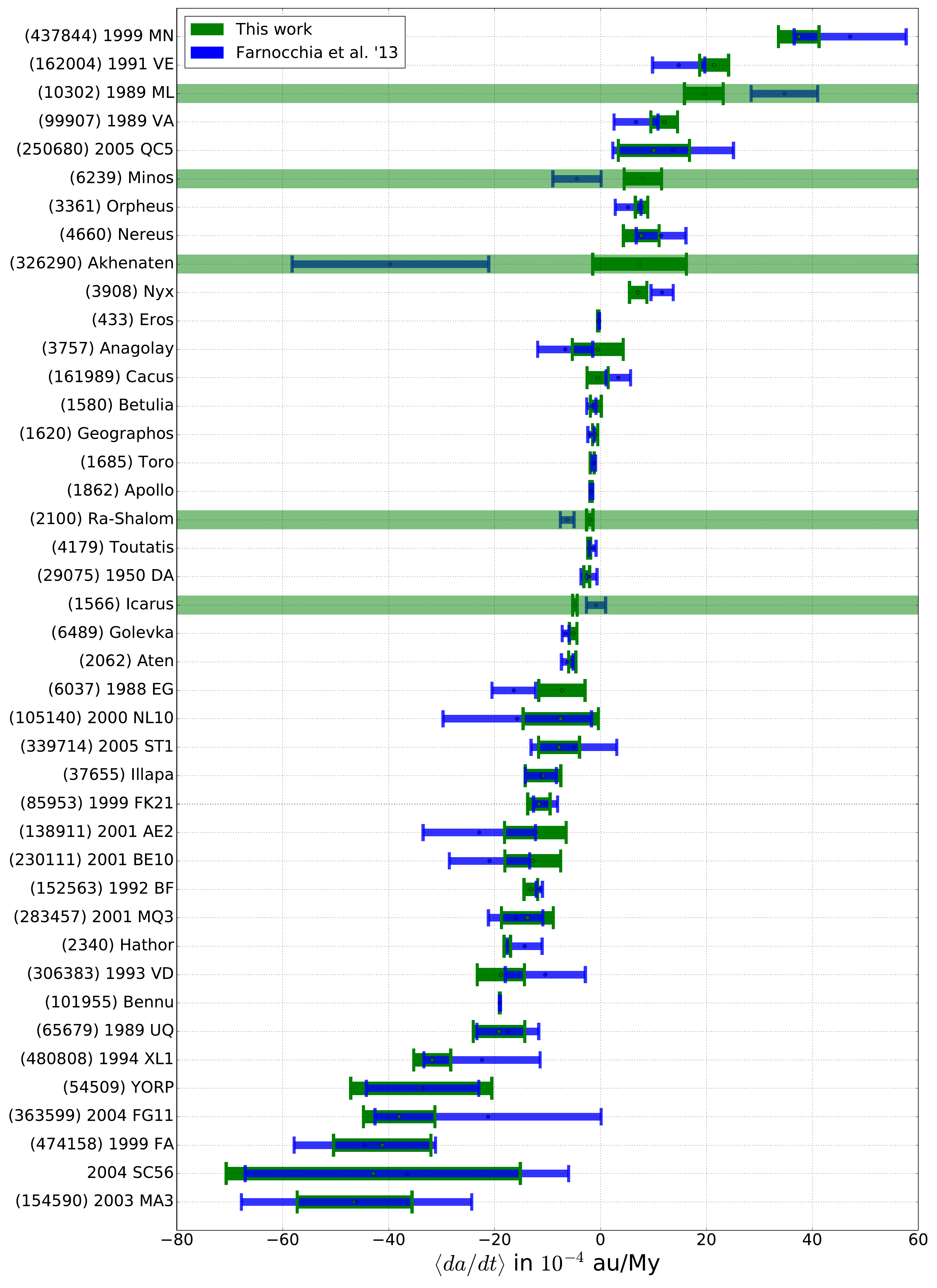}
\caption{A comparison of our Yarkovsky detections (green) and those determined by
  \citet{farn13} (blue), when we used all available data.
  Measurements that disagreed (i.e., $z_i > 2$, Section~\ref{sec:comparison})
  are highlighted in
  green.
  Some objects
  ((483656) 2005 ES70, 2003 XV, 2004 BG41, 2007 PB8, 2009 BD) were not
  included in this plot for display purposes, but all of them had $z_i <= 2$.
  Objects are ranked from most positive to most negative Yarkovsky drift rate.
}
\label{fig:farn}
\end{figure*}

\section{ Comparison with Farnocchia et al. (2013)}
\label{sec:farnocchiacomp}

\subsubsection{ Using matching observation intervals}

We analyzed the 47 Yarkovsky objects found by \citet{farn13} using
matching observation intervals
(to the nearest calendar year) and compared
(Section~\ref{sec:comparison}) our results with their findings. We
found agreement on all $\dadt$ values (Figure~\ref{fig:farnmatch}).

We found that \replaced{four}{five} objects -- \replaced{(105140) 2000
  NL10}{(1580) Betulia}, \added{(3757) Anagolay}, (326290) Akhenaten,
\replaced{(339714) 2005 ST1}{(161989) Cacus}, and 2003 XV -- that were
considered to be detections by \replaced{Farnocchia}{\citet{farn13}}
did not pass our detection threshold
(Section~\ref{sec:refinement}). However, all \replaced{four}{five} of
these discrepant objects are listed in Tables 3 and 4 of
\citet{farn13}, indicating that they are either ``less reliable''
detections or have low S/N values.

\subsection{ Using all available data}

When using all available data, we found relatively good agreement
(Figure~\ref{fig:farn}).  However, we found \replaced{three}{five}
objects --- \added{(1566) Icarus,} (2100) Ra-Shalom, (6239) Minos,
\added{(10302) 1989 ML, and} (326290) Akhenaten --- for which our
drift rates do not match those of \citet{farn13}.  We discuss these
special cases in Section~\ref{sec:farnnomatch}.

\section{ Yarkovsky efficiency distribution }
\label{sec:xi}

Equations (\ref{eqn:yarkfundamental}) and (\ref{eqn:yarkfinal}) provide a
mechanism to interpret the drift in semi-major axis $\dadt$ in terms
of physical parameters of the measured object.  In particular, $\dadt$
can be
described in terms of the Yarkovsky efficiency, $\xi$, where $0 < \xi
< 1$. However, the relationship between $\dadt$ and $\xi$ depends on
density and diameter, and thus determination of $\xi$ requires
estimation of these physical parameters.

Diameters were extracted from the Small Body Database (SBDB)
\citep[][see also Section~\ref{sec:diameter}]{sbdb}.  Densities were
\added{extracted from the SBDB, if available, or } assigned according
to \deleted{SMASS II} taxonomic types, which we also extracted from
the SBDB, using the mean \replaced{densities}{values} reported by
\citet{carry2012}.  Objects of unknown \added{density or} taxonomic
type were assigned a density equal to the \replaced{mean}{median}
density (2470 kg/m$^3$) for the objects in our sample with known type.

\begin{figure}[h!!]
  \centering
  \includegraphics[width=1.0\columnwidth]{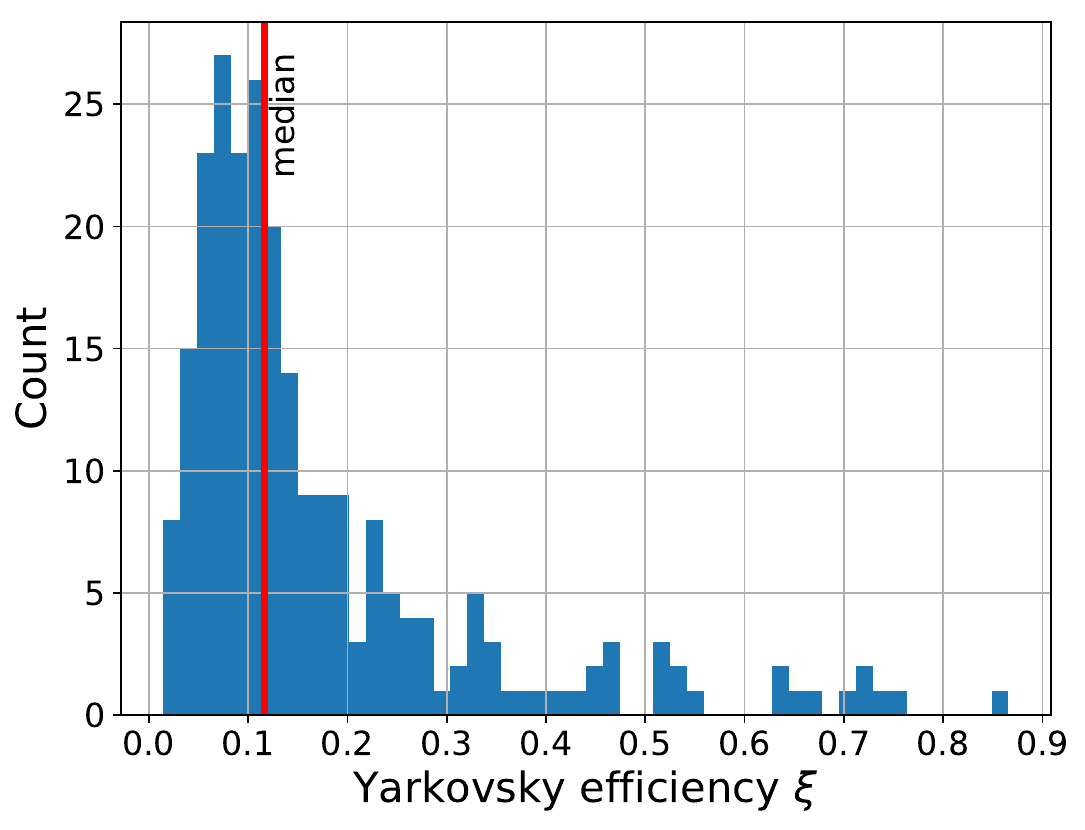}
\caption{ The distribution of Yarkovsky efficiencies $\xi$ measured
  with our sample of \replaced{159}{247} objects.  Diameter and
  density assumptions are described in the text.  The median
  efficiency, $\xi=0.12$, is shown with a red vertical line.  For
  clarity, we did not plot \replaced{two}{four} objects with unphysical
  ($>$1) efficiencies (Section \ref{sec-unphysical}).}
\label{fig:xihist}
\end{figure}

We analyzed the distribution of $\xi$ values and found a median
Yarkovsky efficiency of \replaced{$\xi = 0.12_{-0.06}^{+0.17}$}{$\xi = 0.12_{-0.06}^{+0.16}$}
(Figure~\ref{fig:xihist}).  Note that a bias in this estimate stems
from our inability to
report near-zero drift rates as Yarkovsky detections.
Therefore, the true distribution of efficiencies is presumably
shifted toward lower values than presented here.

Several objects exhibit Yarkovsky efficiencies that substantially
exceed the median value of $\xi=0.12$.  For these objects, the
nongravitational influence, if real, may be unrelated to Yarkovsky
(e.g., sublimation).  It is also possible that some of the
high-efficiency detections are fictitious (e.g., faulty astrometry).
For these reasons, we added a cautionary flag to \replaced{a
  dozen}{20} objects with Yarkovsky efficiencies above 0.5 in
Table~\ref{tbl:data}.  We discuss \replaced{two}{the} unphysical
detections in Section \ref{sec-unphysical}.

\section{ Spin orientation distribution }
\label{sec:ratio}

\citet{laspina2004} provided an estimate
of the ratio of retrograde to prograde
rotators ($N_{R}/N_{P} = 2^{+1}_{-0.7}$) in the NEA population from a
survey of spin
vectors.

Measurements of the Yarkovsky drift rate can also be used to infer $N_{R}/N_{P}$,
because objects with a positive $\dadt$ are almost certain to be prograde rotators,
while objects with a negative $\dadt$ are almost certain to be retrograde
rotators.
This theorized correlation between drift rate and obliquity is borne
out in all cases where both quantities can be estimated \citep{farn13}.

However, given a population of objects with estimated $\dadt$ values, the best
estimate of $N_{R}/N_{P}$ is \textit{not} equal to the ratio $R$ of the number of
objects with negative $\dadt$ to the number with positive $\dadt$.
A bias occurs because
each estimated $\dadt$ value has an associated uncertainty, and there is thus a
nonzero probability that an object with a measured positive $\dadt$ value in fact
has a negative $\dadt$ value (and vice versa). Because there are more retrograde
rotators than prograde rotators, this process will bias observers toward
measuring a lower observed ratio, $R_O$, than is actually present.

This point can be illustrated with a simple (albeit exaggerated)
analytic example. Consider four objects: $A$, $B$, $C$, and
$D$. Objects $A$, $B$, and $C$ all have \added{observed} $\dadt$
values of $-10 \pm 25 \dadtunit$, while object $D$ has \replaced{a}{an
  observed} $\dadt$ value of $+10 \pm 25 \dadtunit$. In this example,
the true ratio, $R_T$, of the number of objects with negative $\dadt$
to the number of objects with positive $\dadt$ is $R_T=3.0$.  However,
when an observer attempts to measure, for example, $\dadt_A$, there is
a $\sim$34\% chance that the observer will erroneously conclude that
$A$ has a positive $\dadt$ value. In fact, we can calculate the
probabilities associated with each of the five possible ratios that
can be observed (Table~\ref{tbl:probs}) and demonstrate that one is
most likely to observe $R_O=1.0$.  If 10,000 observers independently
took measurements of objects $A$, $B$, $C$, and $D$, a plurality would
conclude that $R_O=1.00$, while a majority would agree that $R_O$ lies
between 0.0 and 1.0 --- even though the true ratio is $R_T=3.0$.

\begin{table}[h]
  \caption{ The probability
    ($P$, rightmost column) of measuring a given ratio ($R_O$) of number of
    objects with $\dadt<0$ ($N_{<0}$) to number of objects with $\dadt>0$
    ($N_{>0}$) for a sample of objects with true ratio $R_T=3.0$
    (Section~\ref{sec:ratio}). The true ratio is not the most likely result for
    an observer to measure.}
  \centering
\setlength{\extrarowheight}{0.07cm}
\begin{tabular}{ |r	| r		| r	| r|}
\hline \hline
$N_{<0}$	& $N_{>0}$     	& $R_O$	& $P$ \\
\hline
4		& 0		& $\infty$& $10\%$ \\ %
3		& 1		& 3.00	  & $34\%$ \\ %
2		& 2		& 1.00	  & $37\%$ \\ %
1		& 3		& 0.33	  & $17\%$ \\ %
0		& 4		& 0.00	  & $ 3\%$ \\ %
\hline
\end{tabular}
\label{tbl:probs}
\end{table}

Our data suggest that out of \replaced{159}{247} objects, \replaced{114}{173} have $\dadt < 0$, for an observed ratio of
\replaced{$R_O=\frac{114}{159-114}=2.53$}{$R_O=\frac{173}{247-173}=2.34$}.  To approximate the true ratio $R_T$, we
assumed that the nominal ratio we measured was the most likely ratio for any
observer to measure. Determining the true ratio is then a matter of simulating a
universe with a set of simulated $\dadt$ values that are consistent with our measured values,
and also yield \replaced{$R_O=2.53$}{$R_O=2.34$}.

\begin{figure}[h]
\centering
\includegraphics[width=1.0\columnwidth]{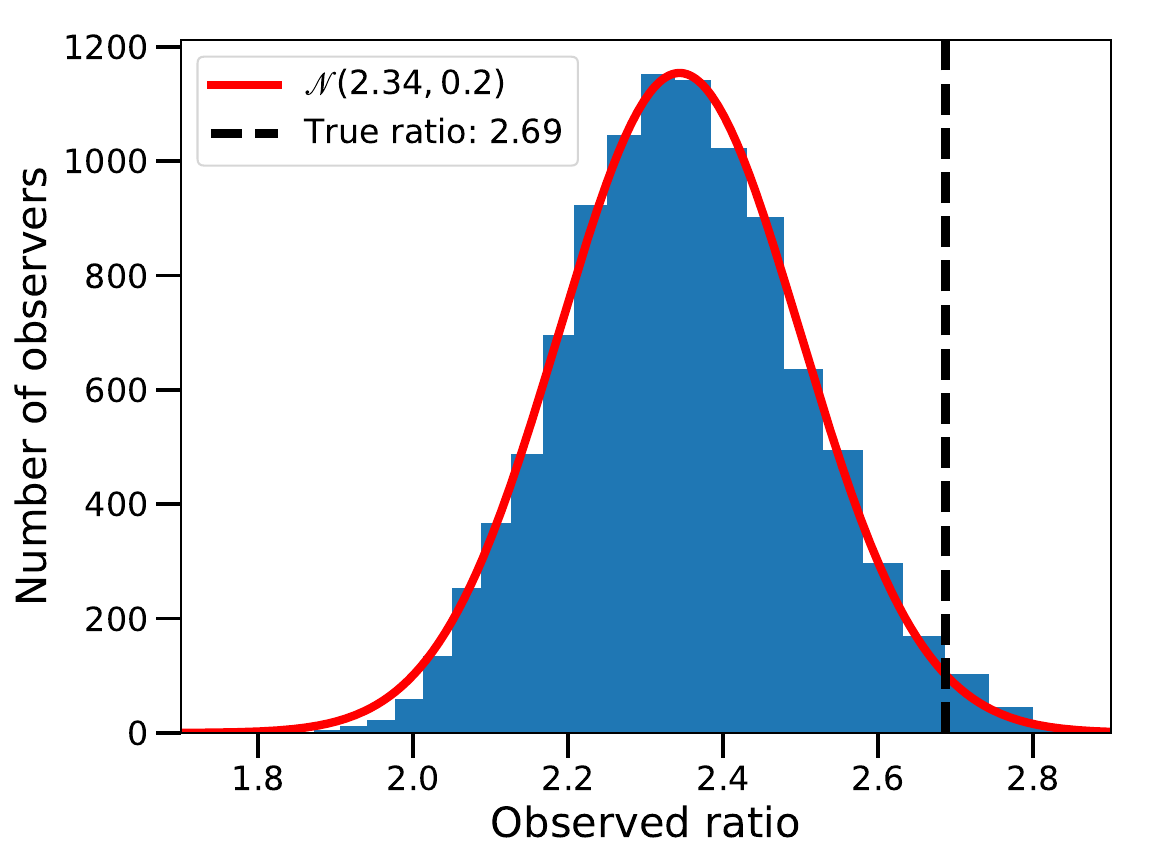}
\caption{ The number of observers measuring a given ratio $R_O$, for $10^4$
independent observers measuring \replaced{159}{247} simulated objects with $\dadt$ values
consistent with what we measured (Section~\ref{sec:ratio}). For a true ratio of \replaced{$R_T=2.9$}{$R_T=2.69$}, most
observers will measure a ratio near \replaced{$R_O=2.53$}{$R_O=2.34$}. This bias must be corrected for
when estimating the ratio of retrograde to prograde rotators from Yarkovsky
observations. }
\label{fig:ratios}
\end{figure}

To find the value of $R_T$ that corresponds to our measured $R_O$ value, we ran
a set of nested Monte Carlo simulations, using the following procedure:
\begin{enumerate}
	\item Create a new `universe', $U_i$.
	\begin{enumerate}
		\item Within $U_i$, generate a set of \replaced{159}{247} $\dadt$ values, pulled
		from distributions consistent with our measurements. This set
		of $\dadt$ values are the true values for the \replaced{159}{247} objects in
		universe $U_i$. Therefore, $R_T$ can be calculated (exactly) for
		this universe.
		\item Simulate what $10^4$ independent observers in universe
		$U_i$ would measure as an observed ratio, $R_O$.
		\item Determine the mean and standard deviation in observed
		ratio ($R_O$ and $\sigma_R$, respectively) in universe $U_i$
		(Figure~\ref{fig:ratios}).
	\end{enumerate}
	\item Repeat step 1 over many ($\sim$$10^3$) universes, and record the
	set of resulting distinct $R_T$ values, and corresponding
	$R_O$, $\sigma_R$ values.
	\item Determine the set of $R_T$ values for which $R_O\pm\sigma_R$
	encompasses our observed ratio of \replaced{$R_O=2.53$}{$R_O=2.34$}.
\end{enumerate}
The resulting simulations suggest that the most likely true ratio for
our observed \replaced{159}{247} objects is
\replaced{$R_T=2.9\pm0.4$}{in the interval $R_T=2.29-2.69$, with a
  strong preference for the upper end of the interval, which
  corresponds to a 72\% fraction of retrograde rotators in our
  sample}.

\replaced{If we wish to relate the ratio of retrograde-to-prograde
  rotators in our data to the corresponding ratio amongst the entire
  population of NEAs}{Because our sample size is limited}, we must also account for
sampling errors, which will further broaden the uncertainties on $R$.
The sampling uncertainty $\sigma_S$ on a measured ratio of $R$ from a
sample of $N$ objects can be calculated directly from the standard
deviation of the binomial distribution and is given by
\begin{equation}
	\sigma_S \approx \sqrt{NR}\times\frac{R+1}{N-R}.
\end{equation}
The sampling uncertainty for $R$ is therefore \replaced{$\sigma_S=0.5$}{$\sigma_S=0.33$}, which suggests a
Yarkovsky-based estimate for the ratio of retrograde to prograde NEAs of 
\replaced{
\begin{equation}
	N_{R}/N_{P} = 2.9\pm0.7
\end{equation}}{
\begin{equation}
  N_{R}/N_{P} = 2.7^{+0.3}_{-0.7}.
  \label{eq:best}
\end{equation}}

The ratio of retrograde to prograde rotators can in principle provide
bounds on the fraction of NEAs that enter near-Earth space through the
$\nu_6$ resonance~\citep{nuge12yark,farn13}.  \replaced{The inference is
complicated by observational bias, namely an overrepresentation of
Atens in the observed sample compared to their expected fraction in a
debiased population.  If we attempt to account for this bias in a
manner similar to that described by \citet{farn13}, we find
  \begin{equation}
N_{R}/N_{P {\rm (debiased)}} = 2.1\pm0.7,
  \end{equation}
  which gives a probability of $\nu_6$ provenance of
  0.35$^{+0.12}_{-0.18}$.  }%
{It is usually assumed that only retrograde rotators can escape
  through the $\nu_6$ resonance and that prograde and retrograde
  rotators have an equal probability of escaping through all other
  routes.  With these assumptions, the fraction of objects that escape
  through the $\nu_6$ resonance can be evaluated as
  $f_{\nu_6}=(N_{R}/N_{P}-1)/(N_{R}/N_{P}+1)$.  Our best estimate of
  the observed ratio (Equation \ref{eq:best}) yields $f_{\nu_6}=46\%$.
  However, the NEA population model of \citet{gran18} suggests that
  our sample contains a factor of 10 overrepresentation of Atens
  compared to their expected 3.5\% fraction in the overall population,
  and therefore a much larger fraction of asteroids originating
  through the $\nu_6$ resonance, $f_{\nu_6} \sim 74\%$.  If the
  population model predictions are correct and the traditional
  assumptions about the sense of rotation of NEAs originating from
  various main belt escape routes are correct, we would expect to
  observe a ratio of retrograde to prograde rotators $N_{R}/N_{P} \sim
  6$, which is more than twice what we actually observe.  Resolution
  of this serious discrepancy may involve one or more of the following
  factors: NEA population model predictions are flawed, assumptions
  about the sense of rotation of NEAs originating from various escape
  routes are incorrect, NEA spin orientations change on timescales
  that are short compared to NEA dynamical lifetimes, or an
  additional, unrecognized bias in our sample exists.  Additional
  estimates of the ratio of retrograde to prograde rotators with more
  stringent detection requirements (lower $p$ values) or a larger
  sample of NEAs (independent of the Yarkovsky sensitivity metric
  $s_Y$) indicate that the observed and true ratio for the entire
  sample do not exceed 2.8.  However, the observed ratios for subsets
  of objects with diameters $>$ 1 km do get larger and closer to the
  $N_R/N_P$ values predicted from the relevant $f_{\nu6}$ estimates.
  This observation suggests that sub-kilometer-size objects are more
  prone to reorientation on short timescales.
  
}

\section{ Yarkovsky effect's diameter dependence }
\label{sec:diameter}

Equations (\ref{eqn:yarkfundamental}) and (\ref{eqn:yarkfinal}) illustrate
the relationship between the magnitude of the Yarkovsky effect and the
affected object's physical parameters. In particular, the theoretical
formulation of this effect predicts a $D^{-1.0}$ dependence.
Verifying this dependence
with our data serves as a check on the theoretical underpinnings
of the effect and also validates our results.

We obtained diameter estimates for objects in our sample from
\replaced{JPL's}{the} SBDB \citep{sbdb}.  For those objects with no
listed diameter, we estimated the diameter from the object's
$H$ magnitude using
\begin{equation}
	D = \frac{10^{-0.2H}}{\sqrt{p_V}}1329 \text{ km},
\label{eqn:diam}
\end{equation}
\replaced{with an assumed geometric albedo, $p_V$, of 0.14 \citep{stuart2004}.}{where the geometric albedo, $p_V$, was extracted from the SBDB, if available, otherwise set to a value of 0.14 \citep{stuart2004}.}
If the uncertainty in diameter was available in the SBDB, we used it,
otherwise we set the uncertainty to a third of the diameter.

Here we note that while the analytical formulation of our Yarkovsky
force model includes parameters that are dependent on the physical
properties of the affected object (Section~\ref{sec:forcemodel}), the
actual fit itself is dependent \textit{only} on dynamics. In other
words, our fits measure  only the overall magnitude of the Yarkovsky
acceleration and are entirely agnostic about physical parameters such
as diameter. Therefore, we can examine the Yarkovsky drift's
dependence on diameter independently from the determination of the
magnitude of the drift itself, and be confident that we are not
committing a \textit{petitio principii}.

\begin{figure}[h]
\centering
\includegraphics[width=1.1\columnwidth]{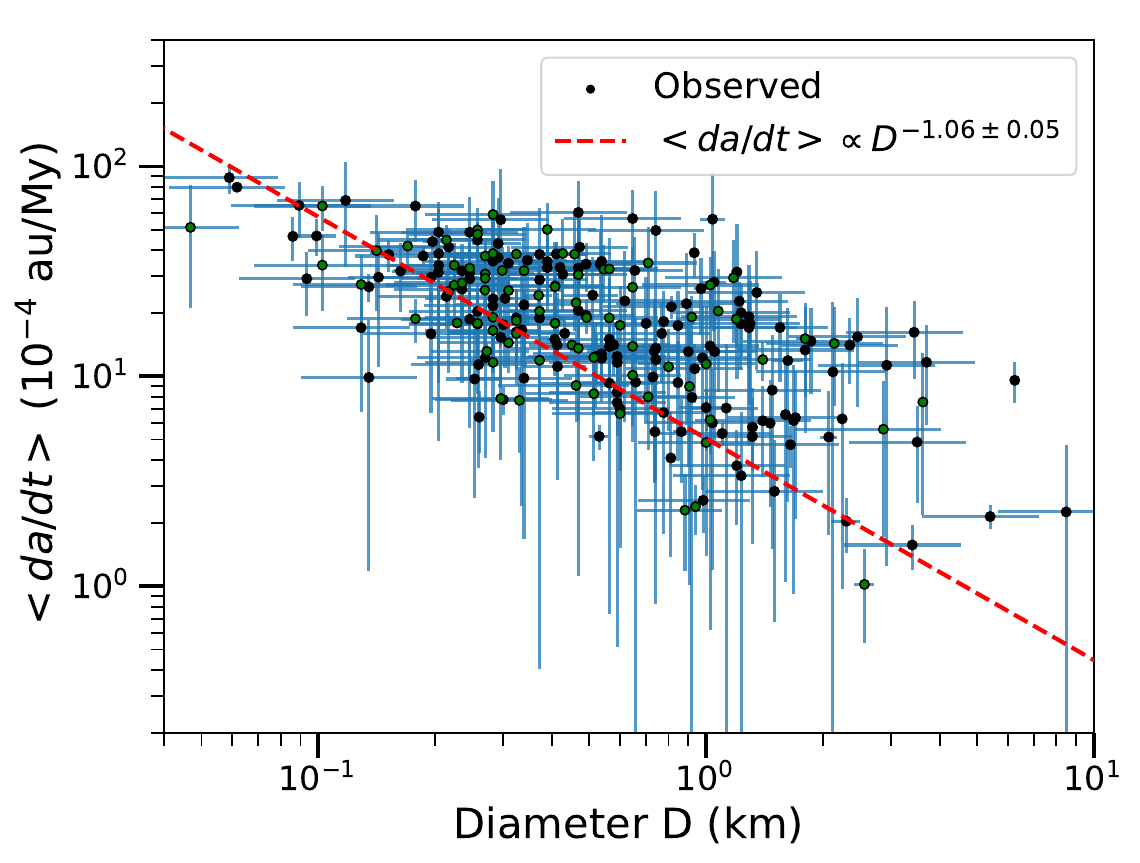}
\caption{ Drift rate $\dadt$ as a function of object diameter,
  $D$\protect\added{, for 244 objects with $\xi<1$}. Diameters were
  either estimated from an $H$ magnitude (black) or extracted from the
  SBDB (green).  Our analysis yields a diameter dependence of
  \replaced{$D^{-1.05 \pm 0.06}$}{$D^{-1.06 \pm 0.05}$}, consistent
  with the theoretical expectation for the Yarkovsky effect of
  $D^{-1.0}$.  }
\label{fig:diam}
\end{figure}

We fit a power law of the form
\begin{equation}
	\dadt = C \times D^p,
\end{equation}
to describe the relationship between the magnitude of the Yarkovsky effect and
the object diameter. We used an Orthogonal Distance Regression (ODR)
\citep{scipy} algorithm to perform this fit, due to the potential errors present in
both the dependent ($\dadt$) and independent ($D$) variables
(Figure~\ref{fig:diam}).
\added{}
The
resulting fit gave a best-fit power-law slope of \replaced{$p=-1.05\pm0.06$}{$p=-1.06\pm0.05$}.
We verified the robustness of this result against the choice of
diameter uncertainties, with values ranging from a fourth to two
thirds of the diameter, and found consistent results. We also verified this result
against different starting conditions on $p$.
\added{In addition, we reprocessed these data using the Python
  software package LINMIX~\citep{linmix,kelly2007}, and found a
  consistent fit of \replaced{$p=-0.96 \pm 0.38$}{$p=-0.93 \pm 0.26$}. We discuss this result further
  in Section~\ref{sec-diam}.}  

\section{ Objects of interest }
\label{sec:objects}

\subsection{ (152563) 1992 BF }

The 1992 BF astrometry includes four optical measurements taken in
1953.  \citet{vokr08} showed that these points suffered from
systematic errors due to faulty catalog debiasing and reanalyzed
these measurements to determine more accurate values. We used these
corrected data \added{and 245 additional observations}, and determined
\replaced{\rate{-13.1}{1.3}}{\rate{-13.2}{1.3}}, which has a z-score
of \replaced{1.62}{1.69} with respect to \citet{vokr08}'s
determination\added{(\rate{-10.7}{0.7})}. The 1992 BF Yarkovsky drift
was also measured with these points from 1953 \added{either
  uncorrected, which yielded \rate{-14.0}{1.3}, or} discarded, which
yielded \replaced{\rate{-14.0}{2.4}}{\rate{-14.1}{2.3}}.

\subsection{ 2009 BD }

\citet{farn13} found a drift rate for 2009 BD of \rate{-493.4}{58.8}.
Following the work of \citet{micheli2012}, \citet{farn13} also fit for
a Solar Radiation Pressure (SRP) model for this object -- which
introduces a radial acceleration as a function of Area-to-Mass Ratio
(AMR) -- and found AMR=$(2.72\pm0.39)\times10^{-4}$ m$^2$/kg.
\citet{micheli2012} found AMR=$(2.97\pm0.33)\times10^{-4}$ m$^2$/kg
with a solution that did not include Yarkovsky.
\added{\citet{mommert2014} performed an
analysis of both spectroscopic and astrometric data for 2009 BD, and found two
possible solutions for AMR, of $1.8^{+0.3}_{-0.2} \times10^{-4}$ m$^2$/kg, or 
$2.2^{+0.4}_{-0.2} \times10^{-4}$ m$^2$/kg.}

We also included an SRP component in our force model for 2009 BD, and
found an area-to-mass ratio of $(2.21\pm0.40)\;\times10^{-4}$ m$^2$/kg,
with a Yarkovsky drift rate of \rate{-497.6}{40.5}. The drastic
improvement in goodness-of-fit when both Yarkovsky and SRP models are
included (Table~\ref{tbl:2009bd}) strongly supports the presence of
these forces. 
\added{This result is consistent with the result found by \citet{farn13} of
\rate{-493.4}{58.8}, as well as
the more modern solution by \citet{vok2015} of \rate{-489}{35}.}
We note that while our uncertainties on the drift rate
appear to be around 20\% better than those of \citet{farn13}, this may
be due to the method by which we fit for $\dadt$, which was performed
as a secondary minimization after fitting for the dynamical state
vector. Therefore, our uncertainties in $\dadt$ do not account for
correlation between parameters and may be an underestimate
because two related, nongravitational effects are present. 

\begin{table}
\caption{ \protect\replaced{Goodness-of-fit}{Sum of squares of residuals} ($\chi^2$) for 2009 BD, using various
  nongravitational dynamical models, with 190 total observations
  prior to outlier rejection. The inclusion of both Yarkovsky forces
  and Solar Radiation Pressure (SRP) yields both a significantly lower
  $\chi^2$, as well as a decrease in the number of outliers,
  $N_{\text{outliers}}$. } \centering \setlength{\extrarowheight}{0.07cm}
\begin{tabular}{ |l	| r		| r    |}
\hline \hline
$Model$		& $\chi^2$     	& $N_{\text{outliers}}$  \\
\hline
Gravity-only	& 109		& 7 \\
Yarkovsky	& 95		& 7 \\
SRP		& 90		& 7 \\
Yarkovsky+SRP	& 75		& 4 \\
\hline
\end{tabular}
\label{tbl:2009bd}
\end{table}

\subsection{ (483656) 2005 ES70 }

The drift in semi-major axis for 2005 ES70 is
\replaced{\rate{-72.8}{5.1}}{\rate{-79.6}{3.1}}. Not only is this a
strong effect, but it is also an unusually strong detection, with a
$p$-value less than $10^{-16}$, and an S/N greater than
\replaced{14}{25}.  \citet{farn13} found \rate{-55.6}{16.7} using
pre-2013 astrometry, which is consistent with our reanalysis of this
object using the same arc (\rate{-54.1}{17.7}).
\added{\citet{vok2015} also found a consistent drift rate of
\rate{-68.9}{7.9}.}

The drop in uncertainty by over a factor of \replaced{three}{five} in
\replaced{four}{six} years is likely due to the increase in data
coverage.  This object has a total of \replaced{132}{172} optical
\added{observations} and \replaced{no radar observations}{a single
  Doppler measurement} since its discovery in 2005.  Of these
\replaced{points}{epochs}, \replaced{48}{83} were measured after 2011 
and were therefore not included in the analysis performed by
\citet{farn13}.  \replaced{This means that the dataset has increased
  in size by over 50\% since 2011, and the arc has grown by
  100\%}{Thus, both the observational arc and the number of observations
  have doubled since 2011}, which \deleted{likely} explains the drop in
uncertainty.

The strength of this effect appears to be anomalous; however, when
we account for this object's small size, we find that its drift rate
is reasonable.  Specifically, the diameter of 2005 ES70 is $\sim$60 m,
as calculated from an $H$ magnitude 23.8 (Equation~\ref{eqn:diam}),
which corresponds to a Yarkovsky efficiency of $\xi=0.06$, assuming a density of
2470 kg/m$^3$.

\subsection{ (1566) Icarus, (66146) 1998 TU3, (66391)~Moshup, (137924) 2000~BD19, (364136)~2006~CJ, (437844) 1999 MN, and (480883) 2001 YE4}
\authorcomment1{Section previously titled: 13.4 (480883) 2001 YE4, (364136) 2006 CJ \& (437844) 1999 MN}

\added{These objects are part of an observational program designed to
  test general relativity and measure the oblateness of the
  Sun~\citep{marg09iau261,verm17gr}.  Incorporating estimates of the
  Yarkovsky orbital drift will be important to improve the reliability
  of the perihelion shift estimates.}

\added{\citet{gree17} presented a detailed analysis of the physical
  and orbital properties of (1566) Icarus, including a Yarkovsky drift
  rate determination and a discussion of the discrepancy with
  \citet{farn13}'s value.  Since 2017, there have been 27 additional
  observations obtained in 2018 and 2019.  We find \rate{-4.84}{0.44},
  which improves upon and confirms \citet{gree17}'s determination of \rate{-4.62}{0.48}.}

\added{(66146) 1998 TU3 is a $\sim$3 km diameter NEA with
\rate{-5.60}{3.9} that cannot be detected with optical astrometry
alone.  The addition of a single range measurement in 2012 magnifies
the difference between the sum of squares of residuals in the zero and
nonzero drift cases, which enables a detection ($p$=0.002).}

\added{(66391) Moshup, also known as 1999~KW4, is one of the best
  characterized binary NEAs (Section \ref{sec:binaries}).  In
  particular, its density is well known \citep[][1974
    kg/m$^3$]{ostr06}.  We found \rate{-5.73}{2.2} and derived
  $\xi=0.043$ on the basis of KW4's measured density.}

\added{(137924) 2000 BD19 is a kilometer-size asteroid whose radar
  detections in December 2007 are notable for being the most distant
  (0.4 au) radar detections of a near-Earth asteroid.  Its Yarkovsky
  drift rate is \rate{-26.2}{10}.}

\deleted{Like 2001 YE4,} 2006 CJ represents a strong Yarkovsky detection with
\replaced{\rate{-38.4}{1.8}}{\rate{-38.2}{1.8}.}\replaced{, and similarly, the}{The} relatively small uncertainty on this
rate is largely due to radar observations. Our analysis includes 11 range and
Doppler measurements of 2006 CJ from 2012 to 2017, and these points reduced the
uncertainty on this detection by $\sim$85\%.

With a drift rate of \rate{37.4}{3.8}, 1999 MN is notable not only for
the high drift rate and S/N, but also for having a semi-major axis
that is increasing rather than decreasing.
\replaced{This object's small semi-major axis ($a=0.67$ au), combined
with a large eccentricity ($e=0.67$), means that this object has a drift that is
over twice as large as that of an asteroid at 1 au with low eccentricity with the same
size and density. The Yarkovsky efficiency for 1999 MN is $\xi=0.05$.}
{Like the other objects in this section, 1999 MN's small semi-major axis and
large eccentricity result in a more pronounced drift rate. While this object's
drift rate is large, the Yarkovsky efficiency for 1999 MN is $\xi=0.05$, well
within the nominal range.}

2001 YE4 has among the largest drift rates in this data set, while
also having amongst the smallest uncertainties, with
\replaced{\rate{-50.1}{0.6}}{\rate{-49.8}{0.7}}. The small uncertainty
is largely explained by the seven radar measurements over three
ranging apparitions --- an analysis of the drift that does not include
these points yields \replaced{\rate{-47.1}{2.1}}{\rate{-48.5}{2.0}},
which means that the radar astrometry reduced the uncertainty by
\replaced{70}{65}\%. The drift rate, while large, corresponds to a Yarkovsky
efficiency of $\xi=0.13$, which is close to the median efficiency for
the objects we analyzed.

\deleted{All three objects are part of an observational program designed to
test general relativity and measure the oblateness of the
Sun\\\citep{marg09iau261,verm17gr}. Their Yarkovsky drift rates will be
taken into account in future analyses.}

\subsection{ (4179) Toutatis }
\label{sec-Toutatis}

(4179) Toutatis is the only object in our sample for which our rate disagreed
with a previous work's result when using similar observation intervals --
namely, our rate of \rate{-2.4}{0.8} has a $z$-score of 2.7 when compared to
\citet{nuge12yark}'s rate of \rate{-5.0}{0.6}. Our rate when using all available
data, \replaced{\rate{-2.7}{0.5}}{\rate{-2.15}{0.3}}, is also not consistent with the previous work's result.

Our rates do agree with \citet{farn13}, who \added{benefited from over
  500 additional observations during the 2012 apparition compared to
  \citet{nuge12yark}'s data set and} found \rate{-1.5}{0.6}.
\citet{farn13} suggest that this object's passage through the Main
Belt may make its orbit particularly sensitive to the number and mass
of gravitational perturbers.

Another curiosity surrounding Toutatis is the drastic change in drift rate that
we found when including radar observations, compared to using only optical
observations -- including radar observations results in an apparent $\sim$80\%
drop in the calculated drift rate.

We found that the difference $\dadt_{o}-\dadt_{r+o}$ between
Toutatis's optical-only drift rate and the radar+optical drift rate is
a strong function of the mass of the 24 Main Belt perturbing objects
included in our force model.  The perturbers included in our
integration account for only $\sim$50\% of the total mass of the Main
Belt.  Artificially increasing the overall mass of these perturbers
brings the $\dadt_{o}$ value into closer agreement with the
$\dadt_{r+o}$ value.
An incomplete dynamical model may therefore explain the discrepancy
between Toutatis's optical-only rate and radar+optical rate.

A final peculiarity about Toutatis is that its orbit can be determined
without any optical astrometry.  We fit our gravity-only and Yarkovsky
models to the \replaced{55}{61} radar measurements obtained over
\replaced{5}{six} apparitions.  The solutions are almost exactly the
same as the solutions that include optical astrometry
(Table~\ref{tbl:radaronly}).  Furthermore, a trajectory fit using only
radar data is consistent with optical data -- the radar-only
trajectory yields a \replaced{goodness-of-fit of
  $\chi_{\text{opt}}^2=1775$ with 11,580 degrees-of-freedom, when
  compared with optical data.}{sum of squares of residuals to optical
  data of $\chi_{\text{opt}}^2=2402$ with 12,070 measurements, i.e.,
  an excellent reduced $\chi_{\text{opt},\nu}^2 \sim$ 0.2.}

These results suggest that the \replaced{55}{61} radar observations
over \replaced{5}{six} apparitions are enough data to obtain a
trajectory that is better than the one inferred from over
\replaced{11,000}{12,000} distinct optical measurements.  \added{A
  similar conclusion was reached with the 55 radar observations
  obtained over the first five apparitions.}

\begin{table}[h]
\caption{Toutatis's orbital elements at epoch 01-JAN-2000 \replaced{00:00:00 UTC}{12:00:00 TDT},
  as determined from radar+optical data,
  and differences in
  orbital element values obtained between
  the optical-only and radar+optical Yarkovsky solutions ($\Delta_{\text{o-only}}$),
  and between
  the radar-only and radar+optical Yarkovsky solutions ($\Delta_{\text{r-only}}$).
  The tiny deviations in the last
  column suggest that optical astrometry is not necessary to determine
  Toutatis's orbit.  Orbital elements include $i$, inclination with respect to J2000.0 ecliptic frame,
  $\Omega$, longitude of the ascending node, $\omega$, argument of pericenter, and $M$, mean anomaly at epoch.
\protect\authorcomment1{Numerical values in this table have been updated.}}
  \centering
\setlength{\extrarowheight}{0.07cm}  
\begin{tabular}{ |l | r | r | r|}
\hline \hline
Orb. element	& radar+optical  	& $\Delta_{\text{o-only}}$ & $\Delta_{\text{r-only}}$ \\
\hline
$a$ (au)		&  2.51054984474     &   1.4e-09 &	 -1.8e-11 \\
$e$ 		&  0.63428487023     &  -1.5e-09 &	 -1.2e-10 \\
$i$ (deg)		&  0.46970399148     &  -2.5e-07 &	  2.5e-08 \\
$\Omega$ (deg)	&  128.367186601     &  -6.5e-06 &	  3.2e-06 \\
$\omega$ (deg)	&  274.683232468     &   1.5e-06 &	 -3.1e-06 \\
$M$ (deg)		& -76.1727086679     &   1.2e-06 &	 -1.2e-08 \\
\hline
\end{tabular}
\label{tbl:radaronly}
\end{table}

\subsection{ (2100) Ra-Shalom, (5660) 1974~MA, (6239)~Minos, (10302) 1989~ML, and (326290)~Akhenaten}
\label{sec:farnnomatch}
\authorcomment1{Section previously titled: 13.6 (2100) Ra-Shalom, (326290)~Akhenaten, and (6239) Minos}

These objects are those for which we found statistically different
results for the drift rate when comparing between our analysis with
modern data and the analysis performed by \added{\citet{nuge12yark}
  or} \citet{farn13} using pre-2013 data.  Our drift rates do match
\added{\citet{nuge12yark} and} \citet{farn13}'s rates when using the
same observational intervals (Section~\ref{sec:farnocchiacomp}).

We found a drift rate of Ra-Shalom
\replaced{\rate{-2.25}{0.77}}{\rate{-2.04}{0.6}}, while
\added{\citet{nuge12yark} found \rate{-5.45}{1.5} and }\citet{farn13}
found \rate{-6.31}{1.3} using pre-2013 data.  \replaced{264}{A total of 686} new
optical observations have been added since 2013, resulting in a
\replaced{$\sim$20\%}{$\sim$50\%} increase in the size of the data
set.  \deleted{While this is not a very large increase,}
\replaced{the}{The} observations since 2013 also include the longest
continuous set of observations ever taken for Ra-Shalom, of around
five months, or $\sim$1/2 of an orbit (here we define a set of
observations as continuous if there is no period spanning more than
two weeks without at least one measurement within the set).
Characterization of the Yarkovsky effect is aided by greater orbital
coverage -- therefore, we expect this modern set of observations to
provide better constraints for this object than was previously
possible. \added{Three range measurements obtained in 2016 also
  improved the solution.}

\added{\citet{nuge12yark} reported a drift rate for (5660) 1974~MA of
  \rate{-20.1}{10}.  In our analysis, we benefited from 95 additional
  observations between 2012 and 2019, an increase of 70\% in the size
  of the data set.  The increased observational interval reveals that
  3 out of the 6 observations reported in 1974 correspond to the 3
  worst residuals.  The Yarkovsky rates obtained with and without
  these 6 observations disagree, which results in a nondetection
  according to our validation tests.}

\replaced{Finally, for}{For} Minos, we find a \added{positive} rate of
\rate{7.98}{3.54}, \replaced{while}{whereas} \citet{farn13} found \rate{-4.45}{4.57} using
pre-2013 data. The number of observations for this object has
increased by over 50\% since 2011, while the length of the observation
interval has increased by 25\%.  The much larger data set explains our
low $p$-value ($p=10^{-5}$), and the shift in the measured effect. We
also note that \citet{farn13} reported this object as a less confident
detection, with S/N$<$2.

\added{We found \rate{19.5}{4} for (10302) 1989~ML whereas
  \citet{nuge12yark} reported \rate{35.3}{7} and \citet{farn13}
  reported \rate{34.7}{6}.  The number of observations has increased
  by $\sim$200 to 520 compared to the pre-2013 data.  We repeated the
  analysis with the pre-2013 data only.  In retrospect, it would have been
  wise to identify this detection as problematic in both studies 
  because five out of the eight observations from 1989 are among the seven
  worst residuals.}

For Akhenaten, we found a drift rate of
\replaced{\rate{7.38}{8.9}}{\rate{7.35}{8.8}}, while \citet{farn13}
found \rate{-39.7}{18.6} using pre-2013 data. Not only do these rates
differ drastically in both magnitude and direction, but we also do not
consider Akhenaten a Yarkovsky detection ($p=0.11$). There have been
\replaced{fewer than 20}{only 18} new observations of this object
since 2012 (a $\sim$7\% increase).  \added{In addition, even if we
  restrict observations to the pre-2013 data, we are unable to obtain
  a Yarkovsky detection.  There is a small number of measurements
  with large residuals, and they are correctly discarded by our
  outlier rejection algorithm.}

\deleted{A clue for the sudden change in apparent drift rate for Akhenaten can
be found by examining the goodness-of-fit metric using pre-2013 data,
$\chi^2_{\text{old}}$, and comparing with the metric when using all
data, $\chi^2_{\text{new}}$ .  In particular, after outlier rejection,
the pre-2013 fit had $\chi^2_{\text{old}}=171$ for 273 data points,
while the fit using all data had $\chi^2_{\text{new}}=151$ for 287
data points. In other words, with the additional data, $\chi^2$
dropped significantly. This is a strong indicator that the difference
in results between the two fits may be due to outlier rejection --
namely, that a small number of points were found to be faulty
measurements with the addition of new data.  If this were the case,
one would expect these points to fall near the outlier rejection
threshold when fitting using pre-2013 data.  We expect that these
faulty observations may have been responsible for producing a false
Yarkovsky detection.}

\deleted{We indeed found three observations of Akhenaten, taken on the same
night from the same observatory (644 Palomar), that were rejected
from the modern analysis, but avoided rejection in the analysis using
pre-2013 data. The three points had residuals of 2.1$\sigma$,
1.9$\sigma$, and 2.7$\sigma$, respectively.  Other than 8 observations from the
prior night,
they were the only measurements of Akhenaten over a ten year period.
Removing these three points from the
pre-2013 data and refitting resulted in a new goodness-of-fit of
$\chi^2_{\text{old}}=135$ (a $\sim$10\% decrease), and resulted in a
Yarkovsky drift rate of \rate{0.91}{18.28}, which is consistent with a
nondetection.
Temporally isolated observations can have a disproportionate effect on a
calculated drift rate. When these observations are also few in number, they
render the perceived rate particularly susceptible to faulty astrometry.}

\added{These examples illustrate the need to carefully review early
  astrometry, which we expand upon in the next section.  }

\subsection{(1036) Ganymed}
\label{sec-ganymed}

\authorcomment1{Section previously titled: 13.7 (174050) 2002 CC19 and (1036) Ganymed}

\deleted{(174050) 2002 CC19 and (1036) Ganymed are the only two objects in our
data set with unphysical ($\xi>1$) Yarkovsky efficiency, with
efficiencies of $\xi=1.29$ and $\xi=4.12$, respectively.}

\deleted{2002 CC19's high efficiency may be due to an incorrect diameter or
density assessment -- this object's spectral type is not known, so it
was assigned a density of 2470 kg/m$^3$ (Section~\ref{sec:xi}).  If
this object had a lower density, perhaps closer to that typical of
C-types, it would drive the $\xi$ value to realistic levels.}

\replaced{Ganymed, however, is a different story. This object's high
  Yarkovsky efficiency is far too high to be explained by an uncertain
  density. However, the data for Ganymed stand out for several
  reasons.} {(1036) Ganymed is a large ($D \sim 37$ km) object that
  provides a good motivation to implement Yarkovsky drift rate
  validation tests that remove early astrometry.  Our nominal solution
  yields an unphysical ($\xi=4.14$) Yarkovsky efficiency, which is too
  high to be explained by an uncertain density.  Despite a
  $p$-value of $<10^{-16}$, our validation tests identified this
  detection as spurious.}  This object has measurements starting in
1924, and thus has one of the longest observational arcs we
considered. It also has one of the largest sets of observations, 
\replaced{($N=5252$)}{$N_o=6359$}.  \citet{nuge12yark} found \rate{-6.6}{1.5}, consistent
with our\replaced{s}{ value} (\rate{-5.0}{1.3}), and devoted a section in their article
to this anomalous case.  \citet{farn13} determined a drift rate
(\rate{-6.1}{1.6}) consistent with \citet{nuge12yark}'s and ours, but
marked it as a potentially spurious detection, due to the unexpected
strength of the drift rate relative to asteroid Bennu's rate scaled
for diameter.  Both \citet{nuge12yark} and \citet{farn13} suggested
that this detection may be due to older, potentially faulty
measurements introducing a false signal.  \citet{nuge12yark} also
explored the impact of an incorrect size or mass determination.

\begin{figure}[h]
\centering
\includegraphics[width=1.0\columnwidth]{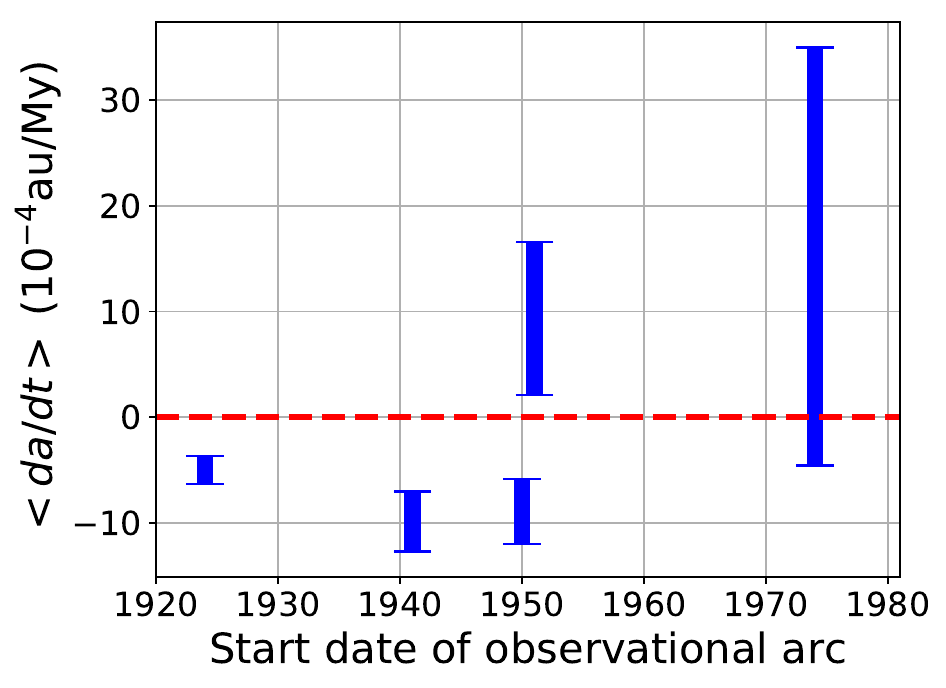}
\caption{ With a drift rate of \rate{-5.0}{1.3}, (1036) Ganymed appears to have
  an unphysical Yarkovsky efficiency of
  $\xi=4.12$.
  We find that if observations
prior to 1950 are discarded, the Yarkovsky effect for this object appears to
abruptly disappear. Ganymed may have an unreliably determined drift rate due to
faulty older astrometry.}
\label{fig:ganymed}
\end{figure}

To examine the possibility that some of the Ganymed astrometry is
faulty, we reran our Yarkovsky determination process after discarding
observations prior to successively later starting dates
(Figure~\ref{fig:ganymed}). We found that the detected drift rate
abruptly disappears if data prior to 1951 are discarded. This fact,
combined with the unphysically large Yarkovsky efficiency
\replaced{required for Ganymed to have a drift rate $|\dadt| > 1.5
  \dadtunit$,}{implied by the large $\dadt$,} leads us to believe
that \replaced{this object represents either a false Yarkovsky
  detection, or a drift rate that}{this object's drift rate} has been
artificially magnified by poor, early astrometry.

\added{
\subsection{(7888) 1993 UC, (217628) Lugh, (461852)~2006~GY2} 
\label{sec-unphysical}

Three objects in our sample appear to have unphysical Yarkovsky
efficiencies, i.e., $\xi>1$.  In this situation, the change in orbital
energy per unit time exceeds the solar power intercepted by an
idealized perfect absorber.  If the diameter, density, and drift rate
are correct, then $\xi>1$ indicates that an additional force is
present, such as that provided by the ejection of dust or volatiles.
However, the diameters, densities, and spectral types of these objects
are unknown or poorly known.

(7888) 1993 UC is a binary asteroid with \rate{-37.79}{26.5} and
$\xi$=1.3.  The fit residuals over a 20-year arc seem reasonable.  The
large error bars allow for $\xi<0.4$.  In the absence of other
information, we assigned a density of 2470 kg/m$^3$ in our
calculations.  If the asteroid's density were in fact closer to 1000
kg/m$^3$, then $\xi$ could be as low as 0.16, close to the median
value.

(217628) Lugh has \rate{-97.09}{40.0} and $\xi=1.88$.  The fit
residuals of 114 observations obtained between 1989 and 2019 are
unremarkable.  In addition, there are six observations in 1960 with
good residuals, for a total observational interval of 59 years.  The
diameter value (1.4 km) in the SBDB appears to come from an earlier
determination of the absolute magnitude and 0.15 albedo assumption.
With the current H value (16.5) and 0.14 albedo assumption, the
diameter would be closer to 1.8 km and $\xi$ closer to 2.4.  We find
that it would take a combination of factors to bring $\xi$ to a
reasonable value.  For instance, a 0.04 albedo, near-unit density, and
the lower range of the drift rate would bring $\xi$ to 0.29.  While
this combination of factors is possible, it is also possible that Lugh
experiences a truly anomalous drift rate.

(461852)~2006~GY2 is a binary asteroid with \rate{-319.6}{157} and
$\xi$=3.47 derived from a 10-year arc.  Fit residuals are
unremarkable.  Here also, it would take a combination of factors to
bring the $\xi$ value below 0.35, suggesting that this object may
experience an anomalous orbital evolution.

}
\subsection{ (99942) Apophis}
\authorcomment1{Section added at reviewer's request.}

Apophis has previously had a Yarkovsky drift detection of
\rate{-23}{13} \citep{vok2015b}. \replaced{Our candidate selection
  algorithm (Section~\ref{sec:selection}) rejected Apophis as a valid
  Yarkovsky candidate}{Apophis was not included in our list of
  candidates because it scores low (0.5) on the Yarkovsky sensitivity
  metric $s_Y$.} \replaced{-- however}{However}, an analysis of this object's astrometry with
our orbit determination software does indeed find a drift rate of
\rate{-25.6}{13.6}, with a $p$-value of \replaced{$2.2e-9$}{$10^{-9}$}. This independently
confirms the previous finding of \citet{vok2015b}.

\deleted{Similarly, our selection algorithm rejected the binary 2000 DP107, but
our analysis of this object's astrometry yields a drift rate of\\
\rate{5.6}{2.3}, $p=1.4e-5$.}

\replaced{These results suggest that our filtering algorithm}{This
  result suggests that our initial screening} can be overly
restrictive by rejecting objects that do have detectable drift
rates. \replaced{We plan to address incidences of false negatives for
  future analyses.}{An obvious solution would be to attempt detections of
  Yarkovsky rates for all NEAs, which is beyond the scope of this
  work.}

\subsection{ Binary and triple asteroids }
\authorcomment1{Section previously titled: 13.8 Binary asteroids}
\label{sec:binaries}

\replaced{Our sample of objects include four confirmed binary
  asteroids -- \deleted{(1862) Apollo,} (136993) 1998 ST49, (363599)
  2004 FG11, and (385186) 1994 AW1.}{Across our data sets, we
  considered a total of 44 numbered binary or triple asteroids.  Among
  these, 18 have $p<0.05$ and 14 passed all our
  detection validations.  These systems are flagged with a $^B$ in Table~\ref{tbl:data}.} Binaries present an opportunity to infer
thermal properties from a Yarkovsky measurement, because tight
constraints can be placed on both mass and obliquity for these
objects~\citep[Section~\ref{sec:interpretxi},][]{marg15AIV}.

\added{Notable systems with Yarkovsky rate determinations include the
  well-characterized 1999 KW4, aka (66391) Moshup
  \citep{ostr06,sche06}, (185851) 2000 DP107 \citep{marg02s,naid15dp},
  and the triple (136617) 1994 CC \citep{broz11,fang11triples}, as well as the
  potential mission target (175706) 1996 FG3 \citep{sche15}.  Other
  notable systems initially passed our detection threshold ($p<0.05$)
  but were subsequently removed because their detections were not
  robust against removal of early observations.  They include (1862)
  Apollo with \rate{-1.78}{0.22}, (65803)
  Didymos~\citep{prav06,fang12spinorbit} with \rate{37.8}{27},
  the triple (153591) 2001 SN263~\citep{fang11triples,beck15} with
  \rate{60.1}{36}, and (285263) 1998 QE2~\citep{spri14} with
  \rate{-59.0}{82}.}

\section{ Discussion}
\label{sec:discussion}

\subsection{ Population-based detection verification }
\label{sec:verification}

We have presented a statistical test that can be used to verify that
a Yarkovsky detection is valid. However, one might still make the
argument that the detections presented herein are merely due to
statistical fluctuations. After all, the Yarkovsky effect often
results in extremely small variations in an orbit. Perhaps the
detections we present are really just a side effect of adding an extra
degree of freedom to the gravity-only dynamical model.

Given the number of objects in our samples, we can address these
concerns by looking for verifications of our detections on a
population level, in addition to object by object.  One such
verification is the correspondence between the measured $\dadt$ versus $D$
inverse relationship, and the relationship predicted by the Yarkovsky
theory (Section~\ref{sec:diameter}). It seems unlikely that a process
that is merely fitting for statistical noise would generate the $1/D$
behavior that we expect \textit{a priori}.

Another population-level analysis considers the distribution of spin
poles of NEAs. We have already discussed how we measured the ratio of
retrograde to prograde rotators in our sample \added{and evidence of
  the agreement between the orbital drift directions and the object's
  obliquities} (Section~\ref{sec:ratio}). We can also use the raw number
of negative $\dadt$ values compared to positive $\dadt$ values to test
the ``statistical noise hypothesis.''  Namely, we can ask the following
question: if our dynamical model, purportedly measuring a
nongravitational force, were instead merely overfitting for
statistical noise, what would be the probability that we would have
measured the number of retrograde rotators that we saw in our sample?
In other words, what is the probability $P$ of achieving a particular
number $m$ (or more) of negatively signed $\dadt$ values in a population
of $N$ objects?

This question can be rephrased in terms of the probability $P$ of of
observing at least $m$ heads after $N$ coin \replaced{flips}{tosses},
for a coin weighted with probability $p$. This can be answered using
the binomial distribution
\begin{equation} \label{eqn:binomial}
	P = 1-B(m-1,N,p).
\end{equation}
In our sample, we have \replaced{$m=114$}{$m=173$} objects with a negative $\dadt$ out of
\replaced{$N=159$}{$N=247$} objects total. To determine $p$, we first assume that the
nongravitational dynamical model \textit{is} in fact overfitting for
noise. In that case, the extraneous parameter would not favor one sign
or another -- in other words, the distribution of $\dadt$ values that
are measured should have a median of 0, which would suggest $p=0.5$.

\replaced{Putting these values into Equation (\ref{eqn:binomial}), we
  find $P=2.2\times10^{-8}$.}{The theoretical probability of observing
  173 heads in 247 tosses of a fair coin is $\sim$10$^{-10}$ (Equation~\ref{eqn:binomial}).  In order
  to avoid making precise statements on the basis of a small sample,
  we report this probability as $P\ll0.0001$.}  \replaced{In other
  words, if}{If} the model were merely measuring \added{unbiased} statistical noise,
the odds of finding the ratio of negatively signed to
positively signed drift rates observed in our data set (or a ratio
more extreme) is \replaced{approximately 1 in 46 million}{much less
  than 1 in $10^4$}.  This extremely low value provides an {\em ab
  absurdo} refutation of the hypothesis that we are fitting for noise.
Note that this probability was calculated with minimal assumptions
about the nature of the underlying statistical noise -- we need only
assume some distribution with a median of $\dadt=0$.

\subsection{ The viability of Yarkovsky measurements }

For those objects with previous Yarkovsky detections, we have compared
results from two previous works (namely, \citet{nuge12yark} and
\citet{farn13}) and found excellent agreement
(Section~\ref{sec:comparison}). The general strength and consistency
of the agreement when using roughly similar observation intervals
(where we found disagreement on drift rates for only a single object)
serve as a validation of the methods employed by all three
groups. The agreement when we used all data available to us (where we
found disagreement on drift rates for only \replaced{five}{six}
objects) speaks to the viability of measuring this small effect from
astrometric measurements, because the measured rates are
\added{generally} stable, even with the addition of new data.
\added{However, in 5--10\% of reanalysis cases, we found substantial
  differences with previously reported detections.}

Among this work and the two previous studies, at least three different
orbital integration packages were used to perform the analyses,
indicating robustness of the results against numerical implementations.

\added{\subsection{ Using the Yarkovsky efficiency $\xi$ to detect
    anomalous rates }

The Yarkovsky efficiency is a fundamental measure of the ability of
asteroids to convert solar energy into orbital energy.  The only
quantities required to evaluate $\xi$ are the Yarkovsky drift, which
can be obtained with a sufficiently long arc of astrometric
measurements, the asteroid's diameter, and the asteroid's density
(Equation \ref{eqn:yarkfundamental}).  We evaluated $\xi$ for a large
sample of asteroids and found a relatively narrow distribution with a
median value of $\xi=0.12$.  Objects with $\xi$ values larger than
$\sim$0.5 are likely anomalous, where the anomaly may be caused by
either grossly incorrect density or diameter values, questionable
astrometry, or orbital influences beyond the Yarkovsky effect, such as
those experienced by active asteroids.  These objects are noteworthy
and deserve further investigation.

When it comes to the identification of anomalous Yarkovsky drift
rates, we find the use of $\xi$ more compelling than \citet{farn13}'s
use of Bennu's drift rate scaled for diameter, for two reasons.
First, scaling by diameter is equivalent to assuming that all
asteroids have a density identical to that of Bennu.  Second,
comparison to Bennu's rate may introduce errors because Bennu's
Yarkovsky efficiency ($\xi=0.785$) does not appear to be
representative.}

\subsection{ Using the Yarkovsky efficiency $\xi$ to provide insights into NEA thermal properties}
\label{sec:interpretxi}
\authorcomment1{Section previously titled: Interpreting $\xi$ }

We have found that within our sample of objects, typical Yarkovsky
efficiencies lie between \replaced{$0.06-0.29$}{0.06 and 0.27} (Section~\ref{sec:xi}).  An
in-depth interpretation of these values would require a full thermal
model of each object. However, we can still provide insights by making
the simplifying assumption that all absorbed photons are reemitted
equatorially. Then, 
the $\xi$ values can be interpreted relative to the
obliquity and thermal properties of the object
in one of three ways:
\begin{enumerate}
	\item If all the reradiated photons were emitted at the same phase lag of
	$\phi=\pm90\degs$, then the obliquity would be $\gamma \sim \arccos{\xi}$. With
	these assumptions, our typical $\xi$ values suggest a range of obliquities 
	\replaced{73}{74}$\degs<\gamma<87\degs$ or $93\degs<\gamma<$\replaced{107}{106}$\degs$.
	\item If the obliquity, $\gamma$, were $0\degs$ or $180\degs$, and all the reradiated
	photons were emitted at the same phase lag, then the phase lag would be
	$|\phi| \sim \arcsin{\xi}$. With these assumptions, typical efficiencies of
	\replaced{$0.06<\xi<0.29$}{$0.06<\xi<0.27$}
	imply phase lags of $3\degs<|\phi|<$\replaced{17}{16}$\degs$.
	\item If the obliquity were $\gamma=0\degs$ or $\gamma=180\degs$, and
	the phase lag were $\phi=\pm90\degs$, then
	$\xi$ could be interpreted as a measure of the distribution of photons
	that are emitted around $\phi$.
\end{enumerate}

Item (1) seems unlikely, given that we expect most of these objects to
have obliquities near $0\degs$ or $180\degs$ -- \citet{hanus2013}
found that among a sample of 38 NEAs, more than 70\% had
$\gamma < 30\degs$ or $\gamma > 150\degs$.
Item (2) is more palatable, and its applicability is protected by 
the cosine function's slow drop-off,
which means that assuming very high or very low spin pole latitudes
will introduce errors of less than 10\% for those objects with
$\gamma < 30\degs$ or $\gamma > 150\degs$.

\citet{rubincam1995} derived an expression for phase lag as a function of the thermal
inertia $\Gamma$ of a body rotating at frequency $\nu$, and found
\begin{equation} \label{eqn:phaselag}
	\phi=\text{arctan} \left ( \frac{\Gamma \sqrt{ \nu }}{ \Gamma \sqrt{\nu}+\sqrt{32}\sigma T_0^3}  \right ),
\end{equation}
where $\sigma$ is the Stefan-Boltzmann constant, and $T_0$ is the temperature of
the body when it is at a distance $a$ from the Sun.

With a typical thermal inertia of $\Gamma=200$ J m$^{-2}$
s$^{-\frac{1}{2}}$ K$^{-1}$ \citep{delbo2007},
Equation (\ref{eqn:phaselag}) 
yields a phase lag of $\phi=8.7\degs$ for a body orbiting at a
distance of 1 au and rotating with a period of 4.5 hours. Assuming
$\gamma=0\degs$ or $\gamma=180\degs$,
our median
Yarkovsky efficiency of $\xi=0.12$ suggests $|\phi|=7\degs$, which is in
good agreement with the phase lag derived from thermal properties.
With a more complete thermal model, it should be possible to relate
any of the differences between these two determinations to the
distribution of reemitted photons (Item (3) above).
\added{We also note that these estimates do not take into account biases that may be
present when translating between measured drift rates and phase lag in the
absence of a full thermal model.}

Better knowledge of $\gamma$ and $\xi$ will yield tighter constraints on thermal
properties of NEAs. In particular, the obliquity and mass of binaries
can be accurately determined through dynamical measurements of the system.
Therefore, binaries with Yarkovsky estimates (Section~\ref{sec:binaries}) will
likely provide the best constraints on thermal properties in the future.

\subsection{Expected diameter dependence}
\label{sec-diam}

\citet{delbo2007} suggested that, due to a dependence between thermal inertia,
$\Gamma$, and diameter, one might expect a flatter $\dadt$ diameter dependence
than predicted by a theory that disregards correlation between these parameters.
In particular, they found that 
\begin{equation} \label{eqn:gammad}
	\Gamma \propto D^{-p},
\end{equation}
where $p$$\sim$$0.4$.

\citet{delbo2007},  citing \citet{vokrouhlicky1999}, wrote
\begin{equation} \label{eqn:vok}
	\dadt \propto D^{-1}\frac{\Theta}{1+\Theta+0.5\Theta^2},
\end{equation}
where 
\begin{equation} \label{eqn:Theta}
  \begin{aligned}
	\Theta &= \frac{\Gamma}{\epsilon\sigma(\sqrt{2}T_{\text{0}})^3}\sqrt{\frac{2\pi}{P}}, \\
  \end{aligned}
\end{equation}
where $\sigma$ is the Stefan-Boltzmann constant, $P$ is the rotation period,
$\epsilon$ is the thermal emissivity, and $T_0$ is the temperature of
the body when it is at a distance $a$ from the Sun.

\citet{delbo2007} suggested that because the asymptotic behavior (i.e., $\Theta \gg 1$)
of Equation (\ref{eqn:vok}) gives 
\begin{equation} \label{eqn:wrong}
	\dadt \propto D^{-1} \Theta^{-1},
\end{equation}
then, by relating Equations (\ref{eqn:gammad}), (\ref{eqn:Theta}), and
(\ref{eqn:wrong}), one would find
\begin{equation}
  \begin{aligned}
	\dadt & \propto D^{-1+p} \propto D^{-0.6}.
  \end{aligned}
\end{equation}

However, few objects yield values for $\Theta$ such that
Equation (\ref{eqn:wrong})'s prerequisite of $\Theta \gg 1$ is
appropriate.  For example, typical objects in our sample have $P=4.5$
hours and $T_0=300$ K.  %
With typical thermal inertias in the range $\Gamma \sim 200-400$ J
m$^{-2}$ s$^{-0.5}$ K$^{-1}$, Equation (\ref{eqn:Theta}) yields
$\Theta \sim 1-2$.  In fact, because Equation (\ref{eqn:vok}) peaks at
$\Theta=1.4$, the slope of the function with respect to $\Theta$ near
$\Theta=1-2$ is nearly 0, which suggests $\dadt \propto D^{-1}\Theta^0
\propto D^{-1}$.

We find \replaced{$\dadt \propto D^{-1.05\pm0.06}$}{$\dadt \propto D^{-1.06\pm0.05}$} (Section~\ref{sec:diameter}), which is
consistent with the nominal theory.

\subsection{ Drift determination and radar ranging }
\label{sec:radarimportance}

While the Yarkovsky effect can be measured for objects with no radar ranging
data, range astrometry aids greatly in improving the accuracy of drift
determination. In particular, the number of distinct radar apparitions with
range data correlates strongly with reduced uncertainty in an object's drift
rate.

\begin{figure}[h]
\centering
\includegraphics[width=1.0\columnwidth]{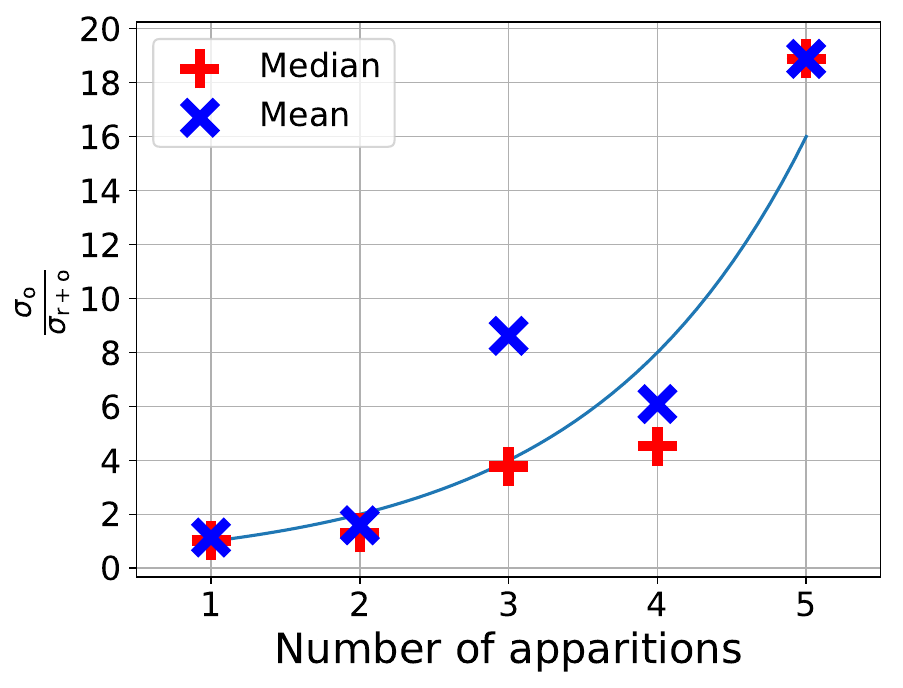}
\caption{ The ratio of the drift uncertainty without radar to that with radar,
$\sigma_{\text{o}}/\sigma_{\text{r+o}}$, as a function of the number of radar
apparitions during which ranging data were taken. The number of objects with
radar range measurements were \replaced{24, 12, 7, and 3}{43, 18, 10, 4, and 1}, for 1, 2, 3, \replaced{and 4}{4, and 5} apparitions,
respectively. }
\label{fig:sigperapp} %
\end{figure}

Of the \replaced{159}{247} objects we analyzed, \replaced{53}{91} had
radar astrometry. Of these, \replaced{46}{76} objects had range
measurements. We examined the improvement in the Yarkovsky
determination -- quantified by
$\sigma_{\text{o}}/\sigma_{\text{r+o}}$, or the ratio of the drift
uncertainty without radar to that with radar -- compared to the number
of radar range apparitions for that object
(Figure~\ref{fig:sigperapp}). \replaced{We found that on average, each
  additional radar range apparition corresponds to an improvement in
  the precision by a factor of $\sim$1.6.}{Although the exact trend is
  obscured by small number statistics, the improvement in precision
  appears to scale roughly as $2^{N_{\rm rad}-1}$, where $N_{\rm rad}$ is the
  number of apparitions with ranging data.}

\section{ Conclusion }
\label{sec:conclusion}

With new astrometry and improved methods, we found a set of \replaced{159}{247} NEAs
with a measurable Yarkovsky drift. We found generally good agreement with previous studies.
Most NEAs exhibit Yarkovsky efficiencies in a relatively small (0--0.2) range.  
We verified the Yarkovsky drift rate's inverse dependence on asteroid size, and 
we estimated the ratio of retrograde to prograde rotators in the NEA
population.  In addition, we provided an estimate of the improvement
in Yarkovsky determinations with the availability of radar data at
multiple apparitions.
Our results provide compelling evidence for the
existence of a nongravitational influence on NEA orbits.

\section{ Acknowledgements }
\label{sec:acknowledgements}

We are grateful for the help and insights provided by Alec Stein
regarding the statistical analyses of our data.  \added{We thank Bill
  Bottke for insights regarding NEA population models and for
  suggesting a possible size dependence to the ratio of retrograde to
  prograde rotators.}

AHG and JLM were funded in part by NASA grant NNX14AM95G and NSF grant
AST-1109772.
\deleted{
Part of the work done here was conducted at Arecibo
Observatory, which is operated by SRI International under a
cooperative agreement with the National Science Foundation
(AST-1100968) and in alliance with Ana G. Méndez-Universidad
Metropolitana (UMET), and the Universities Space Research Association
(USRA).}  The Arecibo Planetary Radar Program is supported by the
National Aeronautics and Space Administration under grant
Nos.\ NNX12AF24G and NNX13AQ46G issued through the Near-Earth Object
Observations program.
This work was enabled in part by the Mission Operations and Navigation
Toolkit Environment (MONTE).
MONTE is developed at the Jet Propulsion Laboratory, which is operated
by Caltech under contract with NASA.
The material presented in this article represents work
supported in part by NASA under the Science Mission Directorate
Research and Analysis Programs.

\software{MONTE \citep{evan18},
	  IDOS \citep{gree17},
	  LINMIX \citep{linmix,kelly2007},
	  ODR \doi{10.1109/MCSE.2007.58},
	  NumPy \doi{10.1109/MCSE.2011.37},
	  Matplotlib \doi{10.1109/MCSE.2007.55}}

\newpage
\FloatBarrier

\clearpage
\bibliography{yark2019}
\end{document}